\documentclass[%
 reprint,
 amsmath,amssymb,
 aps,superscriptaddress,
prb,nobibnotes
]{revtex4-1}
\usepackage{graphicx}
\usepackage{dcolumn}
\usepackage{float}
\usepackage[mathscr]{euscript}
\usepackage{xcolor}
\usepackage[linkbordercolor=green]{hyperref}
\usepackage{tikz}
\hypersetup{
    colorlinks=true,
    linkcolor=red,
    filecolor=magenta,      
    urlcolor=cyan,
    citecolor = red,
}

\usepackage{amsmath}
\newcommand{\bb}[1]{\boldsymbol{#1}}

\newcommand{\ddd}{\mathscr{D}}
\newcommand{\ud}{\textrm{d}}

\newcommand{\revv}[1]{\textcolor{black}{#1}}

\newcommand{\Umax}{U_{\text{max}}}

\newcommand{\dir}{Figs}

\begin{document}

\preprint{APS/123-QED}

\title{
Strong stretching theory of polydisperse curved polymer brushes}

\author{Marios Giannakou }
\email{mgiannak@uni-mainz.de}
 \affiliation{Institut für Physik, Johannes Gutenberg-Universität Mainz}
 
\author{Oleg V. Borisov}%
\affiliation{%
 Institut des Sciences Analytiques et de Physico-Chimie pour l’Environnement et les Matériaux
}%
\author{Friederike Schmid}
 \email{friederike.schmid@uni-mainz.de}
 \affiliation{Institut für Physik, Johannes Gutenberg-Universität Mainz}

\begin{abstract}

\textbf{ABSTRACT:} We investigate the effect of polydispersity 
on the properties of curved linear brushes in good solvent \revv{and for molten brushes}. 
To this end, we extend the strong stretching theory for polydisperse
brushes to curved geometries and investigate the polymer chain end profiles, bending moduli and other properties 
for experimentally relevant polymer chain length distributions of the Schulz-Zimm type.
We also investigate the properties of End Exclusion Zones (EEZ) that 
may appear in convex geometries under certain conditions, and 
show that their position in the brush can be engineered by careful 
selection of the polymer length distribution.
Lastly, we propose a method to engineer chain end profiles by engineering the polymer length distribution.
\end{abstract}
\maketitle
\section{\label{sec:Intro}Introduction}

Tethering polymers onto a substrate is an efficient and versatile strategy to engineer surfaces with adjustable and potentially responsive characteristics. The resultant structures, known as polymer brushes, have been the subject of intensive exploration for several decades. These brushes find applications in diverse fields, including protein immobilization and isolation \cite{protein1,protein2,keating2016polymer}, water filtration enhancement, anti-fouling\cite{keating2016polymer,madhura2018membrane}, bacterial adhesion, surface wettability tuning\cite{azzaroni2012polymer,glinel2009antibacterial,kobayashi2012wettability}, colloid stabilization\cite{zhulina1990theory}, as well as in designing responsive colloids\cite{motornov2007stimuli} and soft surfaces\cite{azzaroni2012polymer,qi2018tuning,qi2015stimuli,qi2020using}.

Theoretical frameworks for understanding polymer brushes have evolved since the 1970s, starting with the seminal work by de Gennes
\cite{de1976scaling} and Alexander \cite{alexander1977adsorption}.
Sophisticated mean-field approaches such as the self-consistent field
(SCF) theory \cite{cosgrove1987configuration} and the analytical
strong-stretching theory (SST) \cite{Milner1988, zhulina1990theory} have facilitated comprehensive investigations into brush properties \cite{binder2012polymer, suo2014self-consistent, yang2015regulating, lebedeva2017dendron, artola2018theory, mikhailov2020brushes}, including monomer density and chain end distribution. Furthermore, such theories are
also used to describe the molecular structure of copolymer-based
micelles, membranes, and bulk mesophases \cite{pickett1996lamellae, matsen2000anomalous, matsen2001testing, heckmann2008strong, matsen2010strong-segregation, tito2010self-assembly, mikhailov2020brushes, mikhailov2021theory, zhulina2022theory, filatov2021microphase, dimitriyev2023medial}.
In theory, it is common to adopt simplifying assumptions, with one prevalent choice being the assumption that brushes consist of polymer chains of uniform length, also known as the monodisperse limit. However, in reality, most polymer brushes exhibit polydispersity, characterized by broad chain length distributions. Despite this, only a handful of papers have delved into the implications of polydispersity. 
An influential semi-analytical extension of SST was devised by Milner, Witten, and Cates in 1989 \cite{Milner1989}.
Subsequent theoretical and simulation studies focusing on polydisperse brushes have highlighted significant variations in their properties compared to their monodisperse counterparts\cite{de2009modeling,qi2016polydisperse}.

A key reason for the prevalence of polydisperse brushes is the limited precision in controlling chain lengths and grafting density afforded by current brush synthesis methods\cite{kim2015self}.
The most precise approach in this regard is the ''grafting to'' method \cite{zdyrko2011polymer,brittain2007structural}, where polymers are synthesized prior to attachment to a surface. This permits a narrow chain length distribution, but comes with the limitation of achieving only relatively low grafting densities. The challenge arises from the difficulty of introducing new chains near the substrate once other polymers have been already introduced.
In the ''grafting from'' method\cite{martinez2016distribution, brittain2007structural}, chains grow monomer by monomer from the solution, originating from initiators attached to the substrate. While this method enables the creation of denser brushes\cite{brittain2007structural}, it comes with the trade-off of less control over the chain length distribution.
Finally, the recently introduced ''grafting through'' \cite{mohammadi2016grafting} adopts a strategy akin to the ''grafting from method'' but employs a porous membrane as a substrate. Monomers are supplied from a solution on the membrane side opposite to the brush. As the monomers approach the initiators from below, shorter chains can grow more readily than longer chains, leading to a more uniform chain length distribution. However, this method has limitations in substrate choice and is less universally applicable compared to other methods.

Here, our focus lies on dense brushes with chain length distributions typical of those resulting from the ''grafting from'' method. In such instances, the chain length distributions exhibit a similar overall form, resembling those of polymers grown in solution. These distributions are often well-described by the Schulz-Zimm distribution\cite{zimm1948apparatus, patil2015demand}.

Recent developments have enabled the design of structures where polymers are attached to spherical nanoparticles\cite{akcora2009anisotropic,
chevigny2011polymer, midya2020structure, bilchak2022understanding}, commonly referred to as "hairy" particles, or onto cylindrical nanopores\cite{adiga2012stimuli, tagliazucchi2012stimuli,
coalson2015polymer, van2019polymer, bilchak2022understanding}. In the former case, polymer brushes have been employed to regulate the distribution of nanoparticles in a solution or to alter the rheological properties of the nanoparticle dispersion. In the latter case, cylindrical nanopores were coated with responsive polymers capable of either closing or opening the nanopore depending on the solvate. It is crucial to note that the polymers attached to these nanopores and particles often possess a radius of gyration that can be of the same order, or even many orders of magnitude larger than the size of the pore or particle. Therefore, it is natural to anticipate significant effects from geometric curvature constraints in this scenario.

The impact of curvature on the structure of polymer brushes has been subjected to theoretical examination for many years by various researchers\cite{ball1991polymers, wijmans1993polymer, dan1992polymers,
manghi2001inwardly,  belyi2004exclusion,lei2015curvature, dimitriyev2021end}.
A notable prediction posits the existence of a ''dead zone'' or ''End Exclusion Zone'' (EEZ), marked by the lack of free end monomers \cite{ball1991polymers,
wijmans1993polymer, belyi2004exclusion, dimitriyev2021end, zhulina2022theory}. However, these investigations have predominantly focused on monodisperse brushes, or, at most, bidisperse \cite{MW_88} brushes. In the latter case, it was found that adding a small amount of longer chains can significantly diminish the bending modulus of molten brushes \cite{MW_88}.

The present authors are only aware of two theoretical studies addressing the conjoined influence of polydispersity and curvature, a Monte Carlo simulation by Dodd and Jayaraman\cite{dodd2012monte}. They observed that polydispersity exerts a pronounced effect on the distribution of chain ends, shifting them closer to the surface eliminating the dead zone in the process. Analytical or semi-analytical mean-field theories 
can help to rationalize and predict such phenomena at low computational cost. In particular, given the historical success of SST and the increasing technological relevance of strongly curved polydisperse brushes, an expansion of the SST to accommodate curved brushes appears highly desirable.

In the present study, we extend the semi-analytical SST framework developed by Milner {\em et al.} for polydisperse brushes\cite{Milner1989}, adapting it to dense brushes on curved surfaces with arbitrary curvature. Our investigation primarily centers on spherical geometries under good solvent conditions, delving into the brush structure across a range of experimentally relevant or theoretically promising chain length distributions. The SST framework yields results consistent with Dodd and Jayaraman's findings. However, depending on the chain length distribution, we find that dead zones are still possible in polydisperse brushes. Moreover, we show that it is even possible to engineer a dead zone such that it appears at some distance from the substrate. Lastly, we show that it is possible to engineer a desired chain end distribution.

The rest of the paper is organized as follows. In section \ref{sec:Model} we introduce the model and briefly go over the foundations, while also providing the analytical extension of polydisperse brushes on curved geometries. In section \ref{sec:results} we present results obtained from the analysis of our model for the Schulz-Zimm distribution  \ref{sec:schulz_results} and explore the effect of polydispersity on the chain end profiles, bending moduli and more. Moreover, in section \ref{sec:eez_results} we explore the emergence of an EEZ in polydisperse brushes and show that it is possible within the framework of SST, for an EEZ to also exist within the brush. Furthermore, in section \ref{sec:chainenddesign} we show that the self-consistent equations in \ref{sec:Model} can be also used to design chain end profiles.

\section{\label{sec:Model}Theoretical Model}

\begin{figure}[t]
    \centering
    \includegraphics[width=0.4\textwidth]{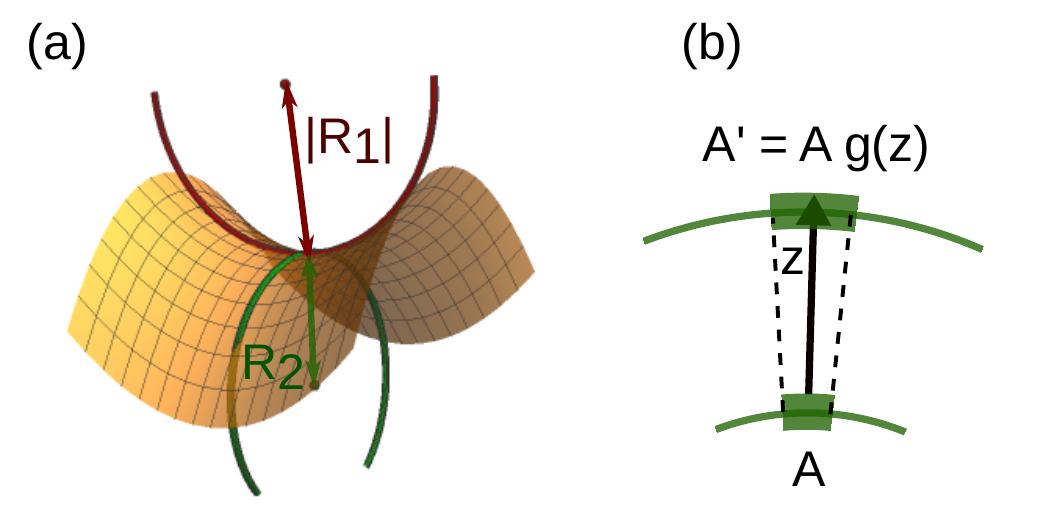}
    \caption{(a) Principle radii of curvature at a point on a 
     curved surface: The largest and smallest radii of tangent 
     circles, where by convention, a radius is negative if the
     curvature turns upwards (as for $R_1$).
     (b) Illustration of the Steiner formula.
     }
    \label{fig:curvature}
\end{figure}

We consider a polymer brush that is grafted upon a curved surface. The
grafting density is given by $\sigma$ and each chain $i$ may have a
different chain length $N_i$, characterizing the number of segments.
The chain lengths are distributed according to a prescribed probability
distribution $P(N)$. The local geometry of the substrate at a given
surface point is characterized by the mean curvature $H =
\frac{1}{2}(\frac{1}{R_1} + \frac{1}{R_2})$ and the Gaussian curvature
$K = 1/(R_1 R_2)$, where $R_1$ and $R_2$ are the radii of curvature
(see Fig.\ \ref{fig:curvature}a). A useful formula regarding 
the geometry is the Steiner formula, which describes the \revv{fraction of areas} between an area element on the substrate and one
which is parallel at a perpendicular distance $z$ 
(see Fig.\ \ref{fig:curvature}b).  It is given by
\begin{equation}\label{eq:steiner}
g(z)=1+2Hz+Kz^{2}.
\end{equation}

We assume that the chains are sufficiently long that they can be
described as Gaussian chains with statistical segment length $a$.
Chains $i$ are thus modelled by continuous curves $\bb{r}_{i}(n)$ with
$n \in [0:N_i]$.  Non-bonded interactions are
described by an energy functional ${\cal V}[\hat{\phi}]$, where
$\hat{\phi}(\bb{r})$ is the number density of monomers given by
\begin{equation}
    \hat{\phi}(\bb{r})
     = \sum_{i}\int_{0}^{N_{i}} \ud n \,  
        \delta (\bb{r}-\bb{r}_{i}(n)).
\end{equation}
Consequently, the Hamiltonian of a patch of unit area on the substrate is
given by
\begin{equation}\label{eq:ham}
\frac{\mathcal{H}[\{\bb{r}_i\}]}{k_{B}T}
  =\sum_{i} \int_{0}^{N_{i}} \ud n \,
     \frac{3}{2 a^{2}}\left| \frac{\ud \bb{r}_{i}(n)}{\ud n}\right|^{2}
     + {\cal V}[\hat{\phi}].
\end{equation}
Here the sum on the r.h.s. runs over all chains in the patch, the first term 
describes the conformational energy of Gaussian chains,
and the term ${\cal V}[\hat{\phi}]$ accounts for nonbonded
interactions between monomers.  In the following, we primarily focus
on the good solvent case, whereupon the excluded volume interaction is
given to first approximation by
\begin{equation}\label{eq:excl}
    {\cal V}[\hat{\phi}]=\frac{w}{2} \int \ud^3 r \, \hat{\phi}(\bb{r})^{2},
\end{equation}
but the theory can also be applied to other types of potentials,
and to incompressible melts (see Appendix \ref{app:free_energy}).

\subsection{Background: SCF and SST approximations}

In the regime of high grafting density, the chains are closely packed
together, and a chain has many more interactions with its neighbouring
chains than it has with itself. In such an instance, density
fluctuations are suppressed and one can justifiably employ a mean-field
approximation, where chains are taken to move independently in the
average field of the surrounding chains.  This leads to the popular
self-consistent field (SCF) theory \cite{Helfand_75}.  The conformations
of chains $i$ of length $N_i$ are independently distributed
according to the distribution functions 
${\cal P}_i[\bb{r}_i] \sim \exp\big(-{\cal S}_i[\bb{r}_i;W] \big)$ 
with the ''action''
\begin{equation}\label{eq:action}
{\cal S}_i[\bb{r}_i;W] 
  = \int_{0}^{N_i} \ud n \, \Big(
     \frac{3}{2 a^{2}}\left| \frac{\ud \bb{r}_i(n)}{\ud n}\right|^{2}
     + W\big(\bb{r}_i(n) \big) \Big),
\end{equation}
where $W(\bb{r}) = \delta {\cal V}/\delta \phi(\bb{r})$, i.e.,
$W(\bb{r}) = w \: \phi$ in the good solvent case. In the incompressible
melt case, $W(\bb{r})$ acts like a constraint field that
adjusts itself such that $\phi(\bb{r})$ is constant
($\phi(\bb{r})\equiv \bar{\phi}$) everywhere.
The distributions ${\cal P}_i[\bb{r}_i]$ can be used to calculate 
the average segment density $\phi$, but they also depends on $\phi$ 
via $W(\bb{r})$, which creates a self-consistent loop. The SCF
Ansatz reduces the complexity of the problem considerably, but the
numerical solution of the SCF equations remains time consuming
for polydisperse brushes especially in the limit of very long chains 
and high grafting densities.

In this limit, however, a further simplification is provided by the
Strong-Stretching approximation or Strong-Stretching theory (SST)
\cite{Milner1989,zhulina1990theory}.
It assumes that \revv{fluctuations of chain conformations are small, 
because the natural scale of  such fluctuations is the radius of gyration, 
while the end-to-end distance of stretched chains may be much larger 
in dense brushes. Thus the ensemble averages over conformations of
chains with a given length $N_i$ are approximated 
by the most probable ones}, i.e., the ones that minimize the actions
${\cal S}_i[\bb{r}_i(n)]$ (Eq.\ \ref{eq:action}). Due to symmetry, the
problem is now reduced to a one dimensional problem for the
paths $z_i(n)$ in the direction $z$ perpendicular to the substrate. 

Following Milner {\em et al.} \cite{Milner1989}, we further define a
''potential'' $U$ as
\begin{equation}\label{eq:uu}
    U(z)= W(0)-W(z) =: \Umax - W(z).
\end{equation}
Minimizing Eq.\ (\ref{eq:action}) with respect to $z_i(n)$ with
the constraint $z_i(0)=0$ then gives the Euler-Lagrange
equation
\begin{equation}\label{eq:eom}
\frac{3}{a^{2}}\frac{\ud^{2}z_{i}}{\ud n^{2}}
 =-\frac{\ud U}{\ud z}\Big|_{z=z_{i}},
\end{equation}
and the boundary condition $(\ud z_i/\ud n)\big|_{N_i} = 0$. The
latter results from the minimization with respect to $z_i(N_i)$ and 
corresponds to the physically reasonable requirement that chain ends are
tensionless\cite{Milner1988}. Furthermore, we can 
identify integration constants
\begin{equation}\label{eq:energy}
E(N_i) = \frac{3}{2 a^{2}} \left(\frac{\ud z_{i}}{\ud n}\right)^{2}+U(z_{i})
\equiv \mbox{const.},
\end{equation}
which depend on the chain length $N_i$, and can be calculated
as $E(N_i) = U\big(z_i(N_i)\big)$ (using $(\ud z_i/\ud n)\big|_{N_i} =
0$). This allows us to derive an expression for the tension 
$\ud z_i/\ud n$ of a chain $i$ of length $N_i$ at the segment $n$
as a function of the potential $U\big(z_i(n)\big)$ and the
integration constant $E(N_i)$:
\begin{equation}\label{eq:tension} 
  \frac{\ud z_i}{\ud n}
    =\sqrt{\frac{2}{3}} \: a \: \sqrt{E(N_i)-U\big(z_i(n)\big)}.
\end{equation}

Next, we make two further physically reasonable assumptions. First,
the tension $|\ud z_i(n)/\ud n|$ drops monotonically from the grafting
point to the chain end.  This implies, by virtue of Eq.\
(\ref{eq:eom}), that $U(z)$ is a strictly monotonous function.
Second, we assume that chains with a larger length have their ends
placed further away from the substrate than their shorter
counterparts.  As a consequence, a monotonically increasing
relationship can be established between the potential $U$, $z(U)$, and
the length $N(U)$ of a chain which has its end at distance $z(U)$ from
the substrate. 

\subsection{\label{sec:Analytical}SST approach for curved brushes }

Based on the above model and approximations, we can derive a set of
self-consistent equations for polydisperse curved brushes.  In the
following, we will summarize these equations.  A detailed
derivation and an algorithm to solve them are presented in the
appendix, \ref{app:sst_derivation} and \ref{app:sst_algorithm}.

Before listing the self-consistent equations, we note that by
inserting $z(U)$, we can express the local monomer density $\phi(z)$
as a function of the potential $U$. Specifically, for brushes 
in good solvent, we have (using Eqs.~(\ref{eq:excl},\ref{eq:uu}))
\begin{equation}
\label{eq:phi_solvent}
\phi(U) = (\Umax - U)/w .
\end{equation}
For melts, the relation is even simpler:
\begin{equation}
 \label{eq:phi_melt}
\phi(U) \equiv \bar{\phi}_{\text{melt}}.
\end{equation}
Furthermore, we define a ''projected monomer density'', $\lambda(U)$, 
such that $\lambda\big(U(z)\big) \: \ud z$ is the amount of monomers 
between $z$ and $z+dz$ per {\em substrate} unit
area.  Another useful quantity is the ''cumulative end density''
$\sigma_{c}(U)$, which is defined such that $\sigma_{c}(U)/\sigma$ is
the fraction of chains with ends located at distances less than $z(U)$
from the substrate. 

In the absence of an EEZ, our self-consistent set 
then consists of four equations: The first allows one
to determine $z(U)$ from a given function $N(U)$,
\begin{equation}
\label{eq:z_U_abel}
z(U) = \sqrt{\frac{2}{3}} \; \frac{a}{\pi} \:
\int_0^U \ud U' \: N(U') \: 
\frac{1}{\sqrt{U-U' }}.
\end{equation}
The second subsequently gives the projected
monomer density $\lambda(U)$ as a function of $z(U)$,
\begin{equation}
\label{eq:l_U_direct}
\lambda(U) = \phi(U) \: g\big(z(U)\big)
\end{equation}
($g(z)$ was introduced in Eq.\ (\ref{eq:steiner})).
The third expresses the cumulative end density 
$\sigma_c(U)$ in terms of $\lambda(U)$,
\begin{equation}
\label{eq:sc_U_abel}
\sigma_c(U) = \sigma - \sqrt{\frac{2}{3}} \; \frac{a}{\pi} \:
\int_U^{\Umax} \ud U' \: \lambda(U') \: 
\frac{1}{\sqrt{U'-U}}.
\end{equation}
Since $\sigma_c(0)=0$, this equation also encompasses an expression
for $\sigma$. Finally, the fourth equation connects $\sigma_c(U)$ 
with the chain length distribution $P(N)$ and can be used to determine
$N(U)$ from $\sigma_c(U)$,
\begin{equation}
\label{eq:N_U_direct}
\frac{\sigma_c(U)}{\sigma} = \int_0^{N(U)} \ud N'\: P(N')
\end{equation}
which closes the self-consistent loop. 
Eq.\ (\ref{eq:l_U_direct}) follows directly from the definition of the
projected monomer density $\lambda(U)$, and Eq.\ (\ref{eq:N_U_direct})
from the definition of the cumulative end density $\sigma_c(U)$,
taking into account that $N(U)$ increases monotonically.  The
derivations of Eqs.\ (\ref{eq:z_U_abel}) and (\ref{eq:sc_U_abel}) are
detailed in Appendix \ref{app:sst_derivation}. 

In the presence of an EEZ, Eq.\ (\ref{eq:N_U_direct}) does not fully
determine the function $N(U)$, because $\sigma_c(U)$ is constant
inside the EEZ region. This introduces an apparent ambiguity. The
reason is that $N(U)$ has no physical meaning inside the EEZ, since
the EEZ does not contain any chain ends by definition.  However, the
ambiguity can be removed if one adds, as a fifth equation in the
self-consistent set, the explicit requirement:
\begin{equation}
\label{eq:dsc_constraint}
\frac{\ud \sigma_c(U)}{\ud U}  
\left\{ \begin{array}{ll} 
> 0 \; & \text{outside EEZs}\\
= 0    & \text{inside EEZs}
\end{array} \right.
\end{equation}
This requirement, together with
Eqs.~(\ref{eq:z_U_abel})--(\ref{eq:N_U_direct}), then uniquely
determines $N(U)$ in the whole range of $U$. We can rationalize the
physical meaning of $N(U)$ inside the EEZ by imagining a modified
chain length distribution that would be identical to the one considered,
but has a vanishingly small population of chains with the appropriate
chain lengths that would fill in the EEZ. Such a brush would have almost
identical properties than the original brush, but $N(U)$ would be 
perfectly defined everywhere.

When can we expect an EEZ?  To answer this question, 
we inspect the following explicit expression for ${\ud \sigma_c(U)}/\ud U$,
\begin{equation}
\label{eq:dsc_U_abel}
  \frac{\ud \sigma_c}{\ud U} = \sqrt{\frac{2}{3}} \frac{a}{\pi}
    \Big( \frac{\lambda(\Umax)}{\sqrt{\Umax - U}}
    - \int_U^{\Umax} \!\!\!\!\!\!\! \ud U'\;
      \frac{\ud \lambda}{\ud U'}
       \frac{1}{\sqrt{U'- U}} \Big) 
\end{equation}
which is derived in Appendix \ref{app:sst_derivation}. The equation
shows that $\ud \sigma_c(U)/\ud U$ can only vanish if there are
regions inside the brush where the projected monomer density,
$\lambda$, increases as a function of $U$ and hence $z$. 
In general, the monomer density $\phi$ decreases monotonically with 
increasing $z$ (see, e.g., Eqs.~(\ref{eq:phi_solvent}) and 
(\ref{eq:phi_melt})). Hence, considering Eq.~(\ref{eq:l_U_direct}),
we see that EEZs can only occur in convex brushes where $g(z)$ 
increases with $z$. A second obvious requirement is the existence
of a gap in the chain length distribution, i.e., $P(N)$ is zero
for a range of chain lengths below the maximum chain length.

In addition to the equations listed above, we also use reorganized
versions of Eqs.~(\ref{eq:z_U_abel})-(\ref{eq:sc_U_abel}) in our
actual algorithm to enforce the requirement (\ref{eq:dsc_constraint}).
They are listed in Appendix \ref{app:sst_derivation}.

Having solved the self-consistent equations, one can use the
solution to calculate the free energy per substrate area $A$
of the brush via
\begin{equation}
\label{eq:free_energy_sst}
\frac{F}{A \; k_B T} = {\cal V}[\phi] 
+ \frac{1}{2} \int_0^{\Umax} \ud U'\; \lambda(U') \: z(U').
\end{equation}
The theoretical background and the derivation of
Eq.~(\ref{eq:free_energy_sst}) are detailed in Appendix
\ref{app:free_energy}. Specifically, the term ${\cal V}[\phi]$
can be expressed as
\begin{equation}
\label{eq:free_energy_solvent}
{\cal V}[\phi] = \frac{1}{2} \int_0^{\Umax}
\ud U' \: \frac{\ud z}{\ud U'} \: \lambda(U') \:
(\Umax - U')
\end{equation}
in the good solvent case, and as ${\cal V}[\phi] = 0$ in the melt
case.  

For completeness we also define one further quantity that we will use
below. One is the projected chain end density $\varepsilon(z)$ which 
is defined as the amount of chain ends between $z$ and $z+dz$ per {\em
substrate} unit area. It can be related to quantities introduced above
in a straightforward manner:
\begin{equation}
\label{eq:chainendrel}
    \varepsilon(z) \; \ud z
     = \frac{\ud  \sigma_{c}(U)}{\ud U} \; \ud U
     = \sigma \; P(N)\:  \ud N.
\end{equation}

In the remainder of this paper, we will \revv{mostly discuss} brushes in good
solvent. We will assume that the chain length distribution $P(N)$ has
an upper bound, and hence the brush has a well-defined height $h$,
above which the monomer density vanishes. In that case, we can
identify $\Umax = U(h)$. We should note, however, that this assumption
is not strictly necessary.

\section{\label{sec:results}Results and Discussion}

We will now present and discuss some results obtained with
the SST framework introduced in the previous section. 
\revv{With few exceptions, we mostly show results for spherical geometries;
the results for cylindrical geometries were qualitatively similar.}

We will first consider brushes with chain lengths distributed according
to the experimentally relevant Schulz-Zimm distribution, 
\revv{which we will characterize with respect to brush height,
chain end density profiles, and curvature elastic moduli.}
Then we will consider other chain length distributions 
and discuss, among other, the emergence of end exclusion zones. 
\revv{Note that we limit our results to values of $\sigma w /a \le1$,
since $w$ has the physical interpretation of an excluded volume,
and $\sigma w/a \approx 1$ thus marks the limit where grafted
monomers on the surface are densely packed.}

\subsection{Polydisperse brushes with Schulz-Zimm distribution}\label{sec:schulz_results}

The Schulz-Zimm distribution is essentially a Gamma distribution with
mean $N_a$ and variance $N_a^2/k$. Experimentally, the variance is
often characterized in terms of the polydispersity index, which 
is defined as the ratio of weight averaged and number averaged chain
length, $N_w/N_a = \langle N^2 \rangle/\langle N \rangle^2 = 1+ 1/k$.

For numerical convenience, we use a truncated version of the Schulz-Zimm distribution:
\begin{equation}\label{schulnew}
P(N) = 
{\cal N}
\begin{cases}

N^{k-1} \exp(- k \: N/N_a)
         ,& \text{if } P(N)> p_{\text{cut}}\\
    0              & \text{otherwise}
\end{cases}    
\end{equation}
where the normalization factor ${\cal N}$ is chosen as to ensure $\int \ud N P(N) = 1$. The maximum cutoff value employed was set to $p_{\text{cut}} = 0.005$.
The truncation effectively establishes both a minimum and maximum chain length. It aids in bypassing numerical issues associated with small numbers of $P(N)$ and also facilitates a straightforward definition of the brush thickness $h$ as the distance from the substrate at which the monomer density vanishes. 

In the following, we present and analyze results derived from the solution of the self-consistent equations relevant to curved polydisperse brushes in spherical and cylindrical geometries. We employ Schulz-Zimm distributions corresponding to polydispersity indices of 1.02, 1.2, and 2. The first index of 1.02 is considered primarily for prospective interests, as achieving such narrow polydispersities is challenging in practical applications. Such precise control of the polymer
size is restricted to special cases such as for example biopolymers or precision
polyamides \cite{hartmann2009precision}. The latter two indices are of greater experimental significance, as they are more easily accessible.


\revv{ Before we proceed to the results, it is important to highlight some
limitations of the Ansatz:  
First, in highly polydisperse brushes, the outermost parts of the longest chains
are expected  to be in the 'mushroom' regime and not in the conditions
assumed by the SST. However, because of the relative small portion of such chains, the expected impact on our results would be minimal. Second, in strongly curved concave geometries, the SST becomes questionable if the width of the brush layer becomes comparable to the curvature radius. If the free ends of polymers that have grafted ends in different substrate regions intermingle, one expects a cross-over from the strong-stretching regime to a concentrated bulk melt, and the theory should be modified accordingly. If the brush width exceeds one of the curvature radii, our SST Ansatz cannot be used at all because the Steiner formula (\ref{eq:steiner}) is no longer valid.}

 \begin{figure}
    \centering
    \includegraphics[width=0.48\textwidth]{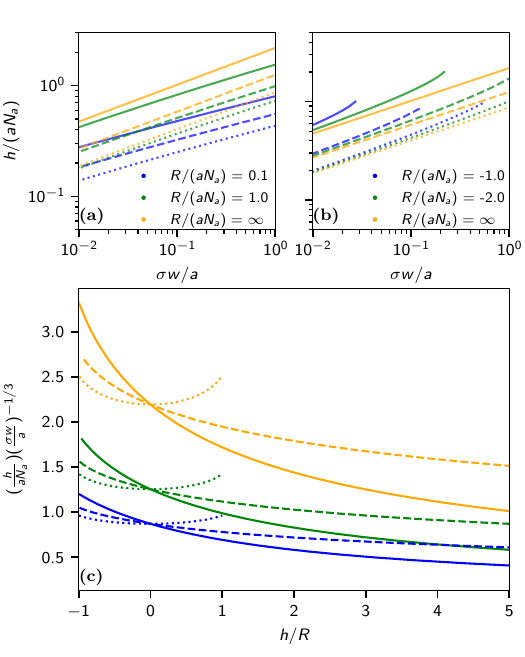}
    \caption{\revv{Scaled brush height $h/(a N_a)$ vs. scaled grafting density $\sigma w/a$ for spherically convex (a) and concave (b)
    geometries and scaled curvature radii $R$ as indicated,} $R=\infty$ represents the planar case.
   The 'dotted', 'dashed' and 'solid' lines, represent polydispersity indices of $N_{w}/N_{a}=$1.02, 1.2 and 2.0 respectively.  
   \revv{Panel (c) shows a universal scaling plot  of the same quantities (see text). Here} the 'solid', 'dashed' \revv{and dotted} lines represent spherical, cylindrical \revv{and saddle} geometries respectively, while 'blue', 'green' and 'orange' colours represent polydispersity indices of $N_{w}/N_{a}=$1.02, 1.2 and 2.0 respectively.}
    \label{fig:three_panel_schulz}
\end{figure}

We first inspect the thickness $h$ of the brushes as a function of grafting density $\sigma$. It is shown for different polydispersity indices in Fig.\ \ref{fig:three_panel_schulz}a (convex brushes) and Fig.\ \ref{fig:three_panel_schulz}b (concave brushes). On curved substrates, the relation between $h$ and $\sigma$ clearly deviates from the straightforward universal dependence in planar brushes, characterized by a scaling exponent of 1/3.

On curved surfaces, such a strict scaling behavior is no longer observed. As the curvature increases, the thickness grows more slowly with $\sigma$ for convex geometries, and more rapidly with $\sigma$ for concave geometries.
The shape of the curves is independent of the polydispersity; however, the polydispersity index has an influence on the prefactor.  
Additionally, according to Eq.~(\ref{eq:steiner}), area elements at a distance $z$ away from the substrate increase or decrease in convex or concave geometries, respectively. 

Consequently, chains encounter more space available at shorter distances in convex curved geometries, leading to a decrease in brush thickness. Conversely, the brush thickness increases with increasing curvature in concave geometries.

We know from the structure of the
SST equations (see Appendix \ref{app:scaled}) that some scaling is still
expected.  Indeed, according to the SST theory, the combination $(h/a
N_a) \: (\sigma w/a)^{-1/3}$ should be a universal function of $h/R$ for
given $P(N)$.  This function is shown in Fig.\ \ref{fig:three_panel_schulz}c \revv{(solid lines)} which summarizes the results in Fig.\ \ref{fig:three_panel_schulz}a and \ref{fig:three_panel_schulz}b. \revv{In addition we have also plotted the corresponding curves for cylinder geometries (with $H=1/2R$, $K=0$) and for saddle like geometries where $H=0$ and $K=-1/R^{2}$. Comparing these curves}
further confirms the claim that the geometry essentially governs the functional dependence of the thickness $h$ on the grafting density $\sigma$, while the polydispersity mainly affects the prefactor.
\revv{While the curves decrease monotonically as a function of $1/R$ in spherical and cylindrical geometries, the curves for saddle points show qualitatively different behaviour, they resemble parabolae centered at zero. This is due to symmetry, as grafting polymers on either side of such a geometry is equivalent.   }

\begin{figure}
    \centering
    \includegraphics[width=0.48\textwidth]{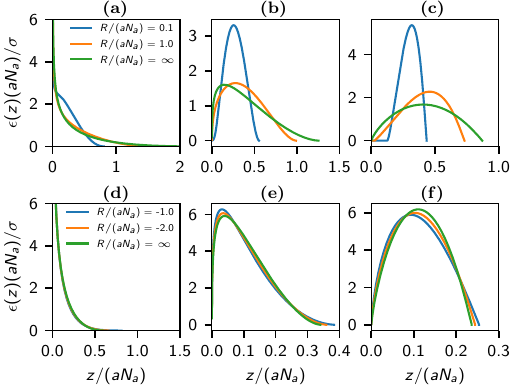}
    \caption{
     Chain end density profiles for brushes in convex (first row \revv{(a)(b)(c)}) and concave (second row \revv{(d)(e)(f)}) spherical geometry. For the convex case, $\sigma w / a$=1, and for the concave case, $\sigma w / a$=0.02. The first \revv{((a)(d))}, second\revv{((b)(e))} and third \revv{((c)(f))} columns correspond to $N_{w}/N_{a}=$2, 1.2,1.02 respectively. The colour in each graph represents the radius $R / (a N_{a})$ of the sphere whose value is displayed in the legend of the first column plot of each row. 
     }
    \label{fig:dis_vs_chain_various_R}
\end{figure}

Continuing our examination, we explore the behavior of the projected chain end distribution, $\varepsilon(z)$, 
for different curvatures,
with $\sigma w /a$ held constant (see Fig.\ \ref{fig:dis_vs_chain_various_R}). The impact of polydispersity on the chain end density profile becomes particularly noticeable in convex geometries. 
Notably, the projected chain end distribution in highly polydisperse brushes with $N_{w}/N_{a}=2.0$ differs qualitatively from those at lower polydispersities: It no longer features a peak at some distance from the substrate, but decays monotonically.
Interestingly, the most probable distance between chain ends and substrate depends on the curvature in a nonmonotonous way for polydispersities of 1.2 and 1.02: It first increases, and then decreases again.
This behavior stems from the concurrent reduction in brush thickness with increasing curvature, and the displacement of chain ends closer to the substrate in convex configurations.

Compared to convex geometries, the curvature has a much smaller influence on the projected chain end distribution in concave geometries, even if the curvature radii are comparable (i.e., $R/a N_a = \pm 1$). Furthermore, as discussed earlier, the SST breaks down for concave geometries if the curvature radius becomes very small. In the range of physically reasonable curvature radii and grafting densities, the projected chain end distributions were not significantly affected by curvature. 

\begin{figure} 
    \centering
    \includegraphics[width=0.48\textwidth]{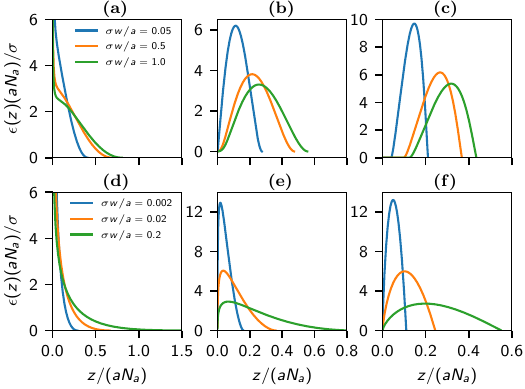}
    \caption{
   Chain end density profiles for brushes in convex (first row \revv{(a)(b)(c)}) and concave (second row \revv{(d)(e)(f)}) spherical geometry. For the convex case is $R / (aN_{a})$= 0.1 and for the concave case $R / (aN_{a})$= -2 . The first \revv{((a)(d))}, second\revv{((b)(e))} and third \revv{((c)(f))} columns correspond to $N_{w}/N_{a}=$2,1.2,1.02 respectively. The colour in each graph represents the value of $\sigma w / a$  whose value is displayed in the legend of the first column plot of each row. 
     } 
    \label{fig:dis_vs_chain_various_sig}
\end{figure}

In both Fig.\ \ref{fig:dis_vs_chain_various_R} and Fig.\ \ref{fig:dis_vs_chain_various_sig}, we see that the narrow distribution $N_{w}/N_{a}=1.02$ leads to an apparent EEZ for high convex curvatures and high $\sigma w /a$. This outcome is a consequence of our truncation of the Schulz-Zimm distribution, which introduces a non-zero minimum chain length in the distribution. Without this truncation, the omitted chains would have populated this region, resulting in a region of highly reduced chain end density rather than a true EEZ. Further insights into EEZs and their characteristics are further explored in the following section.

In Fig.\ \ref{fig:dis_vs_chain_various_sig} the chain end distribution is plotted for a fixed geometry and varying $\sigma w /a$. Here, the effect of competing space can be seen more clearly. As the excluded volume or the grafting density increases, the chains find it preferable to extend their ends further away from the substrate independent of geometry or polydispersity.

In order to further quantify the difference between a given projected chain end density profile $\epsilon(z)$ and the corresponding profile $\epsilon_{\text{pl}}(z)$ for planar brushes with the same chain length distribution $P(N)$ and the same grafting density $\sigma$, we define a profile overlap parameter as follows:
\begin{equation}\label{eq:overlap}
    \text{Overlap }=
    \int_0^1 \ud x \;
 \text{min}\left(\epsilon(x \cdot h) \: \frac{h}{\sigma},
      \epsilon_{\text{pl }}(x \cdot h_{\text{pl}}) \: 
      \frac{h_{\text{pl }}}{\sigma} \right).
\end{equation}

\begin{figure}
    \centering
    \includegraphics[width=0.48\textwidth]{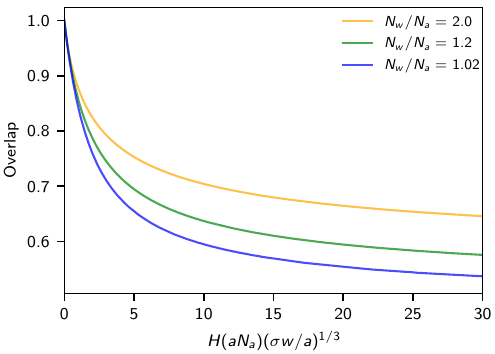}
    \caption{Overlap as defined in Eq.~ (\ref{eq:overlap}) for different polydispersity indices in spherical geometries. }
    \label{fig:ovelap}
\end{figure}

This parameter turns out to be a good measure for quantifying at which curvatures, grafting densities or excluded volume values we expect a significant departure from the planar chain end distribution. As \revv{shown in Fig.\ \ref{fig:ovelap},} the overlap is a decreasing function of the scaled curvature,
$H(aN_{a})(\sigma w /a)^{1/3}$. It reaches a limiting value
for high curvature values. 
Moreover, the overlap increases with increasing polydispersity. This correlation is expected because the necessity for different chains to adjust their position in a progressively crowded environment diminishes if all chains have different length.
Most importantly, the most significant changes to the projected chain end distribution occur in the range of $H(aN_{a})(\sigma w /a)^{1/3}$=0-5. We know from Fig. \ref{fig:three_panel_schulz}c that the ratio $h/(aN_{a})(\sigma w /a)^{-1/3}$ is of order 1-2 for all convex brushes investigated in this work. Hence we conclude that
significant polydispersity effects occur in curvature ranges $ H \: h \sim 0-10$, ranges which are experimentally relevant. 

\begin{figure}
    \centering
    \includegraphics[width=0.48\textwidth]{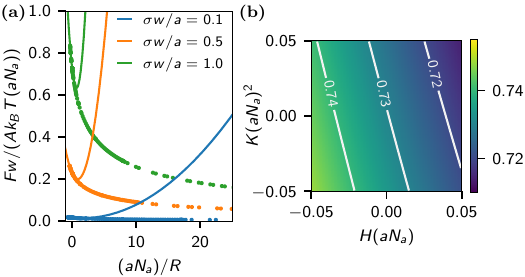}
    \caption{ (a) \revv{Scaled free energy} $Fw/(Ak_{B}T (aN_{a})$ plotted against $(aN_{a})/R$ for a spherical geometry and a polydispersity index of $N_{w}/ N_{a}=$2.0 for different values of $\sigma w /a$. The solid lines \revv{correspond to Eq.~(\ref{eq:helfrich}) with elastic parameters determined as described in the main text.} 
    \revv{(b) 2D plot of the free energy landscape $Fw/(A k_{B}T(aN_{a})$ for a Schulz-Zimm brush with a polydispersity index $N_{w}/N_{a}$=2.0 and $\sigma w /a$=1.0. The free energy data were calculated on a square grid (see text) and interpolated by the fit to Eq.\ (\protect\ref{eq:helfrich}).    
    }  }
    \label{fig:free_energy}
\end{figure}

Lastly, we investigate the \revv{curvature elastic moduli} of polymer brushes as a function of polydispersity. We should note that, generally, polymer brushes experience a force towards outwards bending, since the stretching energy of the brush polymers can be reduced if the brush
surface is curved. This can be seen in \revv{Fig.~\ref{fig:free_energy}a} which shows the free energy as a function of curvature for different
grafting densities and polydispersity index $N_w/N_a=2$. The free energy decreases monotonically with increasing curvature.

The elastic moduli of the surface are defined in the planar limit
$1/R \sim 0$. In this regime of slightly curved surfaces, the expression for the free energy per area is approximated by the so-called Helfrich Hamiltonian \cite{helfrich1973elastic}: 
\begin{equation}
\label{eq:helfrich}
\frac{F}{A} = 
\frac{\kappa}{2}\: (2 H - c_0)^2 + \bar{\kappa} \: K +\revv{\frac{F_{0}}{A}}.
\end{equation}
Here $\kappa$ is the bending modulus, $\bar{\kappa}$ is the Gaussian modulus, \revv{$c_0$ is the so-called spontaneous curvature and $F_{0}$ is 
an surface energy offset.}
To extract the bending and Gaussian moduli, we \revv{have determined the free
energy as a function of $H$ and $K$ for a range of values of $H$ and $K$ in the vicinity of the planar limit, $1/R=0$. Specifically, we used a $7 \times 7$ square grid of evenly space points in the $(K,H)$ plane, in the range  $H(aN_{a})\Umax^{1/2} \in [-0.06:0.06]$ and $K(aN_{a})^{2}\Umax \in [-0.06:0.06]$. To assess the monodisperse limit we used a uniform step-like distribution (see next section) with vanishing width. The results were then fitted to Eq.\ (\ref{eq:helfrich}). An example of a corresponding fitted free energy landscape is shown in Fig. \ref{fig:free_energy}b.}
\begin{figure}
    \centering
    \includegraphics[width=0.48\textwidth]{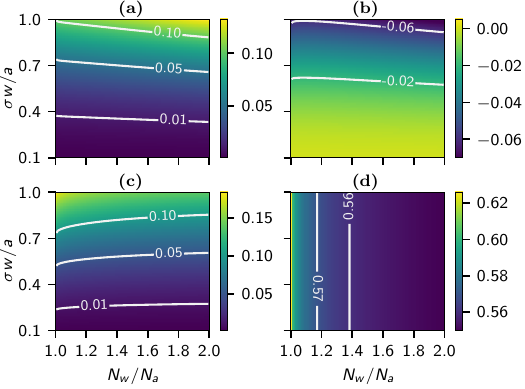}
    \caption{\revv{Elastic curvature moduli (a) $\kappa w/(k_{B}T(aN_{a})^{3})$ \\(b)  $\Bar{\kappa} w/(k_{B}T(aN_{a})^{3})$ (c)  $\kappa c_{0}w/(k_{B}T(aN_{a})^{2})$ (d) $-\Bar{\kappa} / \kappa$ as a function of scaled grafting density $\sigma w/a$ and polydispersity index $N_w/N_a$ for Schulz-Zimm brushes in good solvent. 
    }}
    \label{fig:bending_moduli}
\end{figure}

The Helfrich Hamiltonian was originally proposed for lipid bilayers.
It is important to note that in the brush case, $c_0$ is just a fit parameter and does {\em not} correspond to a preferred curvature, since the true free energy decreases monotonically as a function of curvature as discussed above. However, the parameter $\kappa c_0$
still quantifies the tendency of a planar brush to bend.

In Fig.\ \ref{fig:bending_moduli} ,  the scaled quantities $\kappa w/(k_{B}T(aN_{a})^{3})$,  $\Bar{\kappa} w/(k_{B}T(aN_{a})^{3})$\revv{, $\kappa c_{0}w/(k_{B}T(aN_{a})^{2})$ and $-\Bar{\kappa} / \kappa$} are plotted as a function of 
$N_{w}/N_{a}$ and $\sigma w /a$. As expected, the absolute values of $\kappa w/(k_{B}T(aN_{a})^{3})$ and $\Bar{\kappa} w/(k_{B}T(aN_{a})^{3})$ 
increase with increasing scaled grafting density $\sigma w /a$. 
It is surprising that these two quantities are largely insensitive to the polydispersity\revv{, although a small increase in both moduli is observed with increasing polydispersity. The latter can be explained by the fact that the more polydisperse brushes have a greater portion of small chain population which compete for space near the surface}.

\revv{For comparison, we also show corresponding results for melt brushes
in Fig.~\ref{fig:helfrich_melt}. They were determined in a manner identical to the good solvent case. In molten brushes, polydispersity has the opposite effect on the elastic moduli than in the good solvent case: Polydisperse brushes have lower stiffness than monodisperse brushes, because chains can rearrange more flexibly upon bending of the substrate if they have different length.}
\revv{
\begin{figure}
    \centering
    \includegraphics[width=0.48\textwidth]{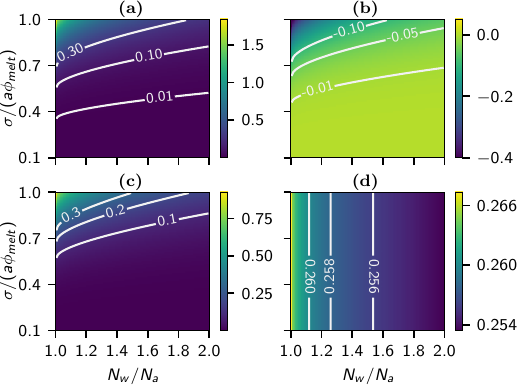}
    \caption{\revv{Similar to Fig.\ \protect{\ref{fig:bending_moduli}} for melt brushes \\(a) $\kappa /(\Bar{\phi}_{\text{melt}}k_{B}T(aN_{a})^{3})$ (b)  $\Bar{\kappa} /(\Bar{\phi}_{\text{melt}}k_{B}T(aN_{a})^{3})$ \\(c)  $\kappa c_{0}/(\Bar{\phi}_{\text{melt}}k_{B}T(aN_{a})^{2})$ (d)$-\Bar{\kappa} / \kappa$. Plotted as a function of $\sigma /a\Bar{\phi}_{\text{melt}}$ and the polydispersity index $N_w/N_a$ for Schulz-Zimm brushes in good solvent.}}
    \label{fig:helfrich_melt}
\end{figure}
}

\revv{Both in the good solvent and the melt case, the ratio $-\Bar{\kappa} / \kappa$ is independent of the grafting density (see Figs.\ \ref{fig:bending_moduli}d and \ref{fig:helfrich_melt}d). This was analytically predicted for monodisperse brushes by Milner and Witten\cite{MW_88}, and it still holds true in the polydisperse case, albeit with a change in the prefactor: The ratio decreases with increasing polydispersity. In the monodisperse limits,
our values agree with analytical predictions by Milner and Witten\cite{MW_88},
$-\bar{\kappa}/{\kappa}\approx 0.61$ for brushes in good solvent, and 
$-\bar{\kappa}/{\kappa}\approx 0.267$ for molten brushes. Furthermore, we 
also inspect the quantity $\kappa c_0$ (Figs.\ \ref{fig:bending_moduli}c and \ref{fig:helfrich_melt}c). It increases with increasing grafting density and decreases with increasing polydispersity, both for brushes in good solvent and for molten brushes. Thus the variety of chain lengths in polydisperse brushes generally reduces the driving force towards bending in the brushes.}
 
\subsection{Polydisperse Brushes with End Exclusion Zones}\label{sec:eez_results}

After addressing brushes characterized by chain lengths distributed according to the experimentally relevant Schulz-Zimm distribution, our attention now shifts to distributions with a truly non-zero minimum chain length, in order to investigate the emergence of an EEZ within polydisperse chain length distributions. For simplicity, we focus on uniform distributions and convex spherical geometries, noting that the findings can be generally extended to other convex geometries as well.

 \begin{figure} 
    \centering
    \includegraphics[width=0.48\textwidth]{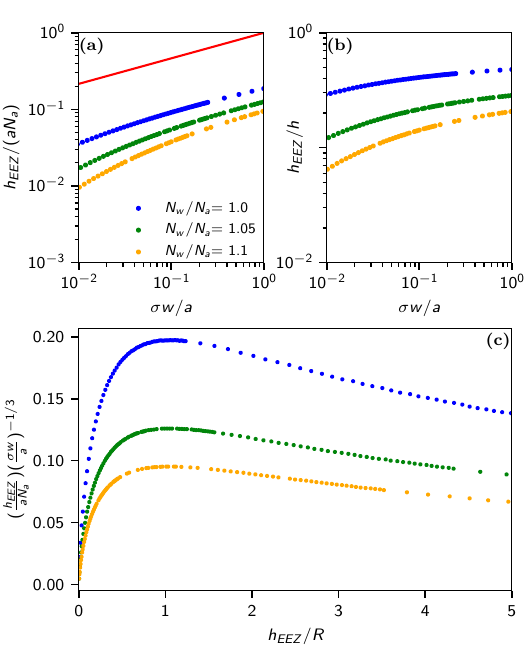}
    \caption{Thickness of exclusion zone $h_{EEZ}/(aN_{a})$ (a) and the ratio $h_{EEZ}/h$ (b) against $\sigma w / a$ for a spherically convex geometry with a radius of $R/(aN_{a})=0.1$ \revv{and polydispersity indices 
    $N_w/N_a$ as  indicated}. The red line in (a) is there for reference and represent a power law with an exponent of 1/3.
    \revv{Panel (c) shows a scaling plot of the same quantities} for spherically convex geometries \revv{for polydispersity indices
    $N_w/N_a = 1$ (blue), $N_w/N_a = 1.05$ (green), and $N_w/N_a = 1.1$ (orange).}}
    \label{fig:sig_vs_eez}
\end{figure}

First we explore the emergence of an EEZ with thickness $h_{eez}$ for a single uniform step-like chain length distribution defined by:
\begin{equation*}
    P(N) = \Biggl\{
    \begin{aligned}
         & \quad \frac{1}{N_{\text{max}}-N_{\text{min}}} \quad ,  N \in [N_{\text{min}},N_{\text{max}}] \\
         & \quad  \quad\quad\quad 0 \quad\quad\quad \, \, \, ,  N \notin [N_{\text{min}},N_{\text{max}}]
    \end{aligned}
\end{equation*}
with non-zero minimum chain length $N_{\text{min}}$, a maximum chain length $N_{\text{max}}$, an average chain length $N_{a}=\frac{1}{2}(N_{\text{min}} + N_{\text{max}})$ and a weight averaged chain length $N_{w}=\frac{1}{3}(N_{\text{min}}^2 + N_{\text{max}}^2+ N_{\text{min}} N_{\text{max}})$. 

We examine three polydispersity indices, with one of them representing the monodisperse limit. Specifically, our focus lies in studying
the EEZ thickness as a function of $\sigma w /a$ at a constant curvature. As anticipated,  Fig.~\ref{fig:sig_vs_eez} shows
an increase in the EEZ thickness with rising values of $\sigma w /a$. Common intuition suggests that a larger value of $N_{\text{min}}$, corresponding to a smaller polydispersity in our setup, results in a lower threshold of $\sigma w /a$ for the emergence of an EEZ. Furthermore, this implies that, for equivalent values of $\sigma w /a$, the thickness of such a region would be larger for larger $N_{\text{min}}$. This is indeed observed in Fig.~
\ref{fig:sig_vs_eez}a. Another noteworthy observation is that the EEZ thickness initially grows rapidly, but for later values of $\sigma w /a$, it appears to increase at the same rate as the thickness of the entire brush, which is demonstrated in Fig.~\ref{fig:sig_vs_eez}b.

Just like in the Schulz-Zimm distribution. we know from the scaling properties of the SST 
equations (Appendix \ref{app:scaled}) that it is possible
to identify hidden scaling laws, i.e., the quantity
$(h_{EEZ}/a N_a) \: (\sigma w/a)^{-1/3}$ can be written as a 
unique function of $h_{EEZ}/R$. This function is shown in Fig.\ \ref{fig:sig_vs_eez}c. As it can be deduced the scaling behaviour is different than the scaling behaviour observed in Fig.\ \ref{fig:three_panel_schulz}c, albeit for higher values of $h_{EEZ}/R$ or $h/R$ their scaling behaviour appears similar.

\begin{figure}
\centering
   \includegraphics[width=0.7\linewidth]{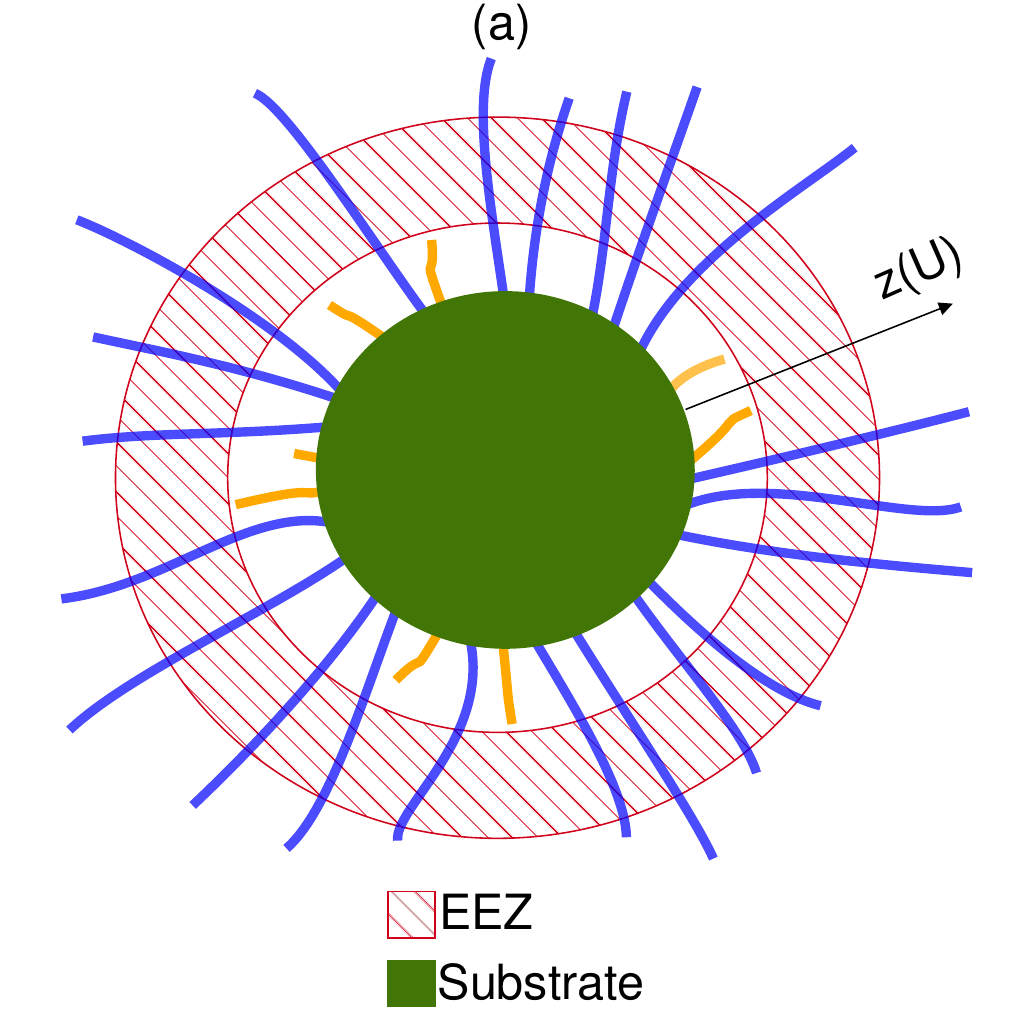}
\\
   \includegraphics[width=1\linewidth]{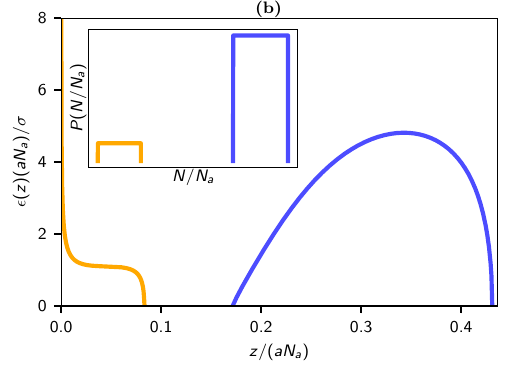}
\caption{Projected chain end density against the distance from the substrate \revv{(b)} for a spherically convex geometry with $\sigma w /a$=1 and $R/(aN_{a})$=0.1. The inset represents the corresponding chain length distribution of the polymer brush. The orange color represent chains with a size below $N_{0}$ while blue represents chains with size above $N_{1}$. A cartoon representation \revv{(a)} of the polymer brush in question is also shown where the color coding is the same as the plots below it.}
\label{fig:eez_middle}
\end{figure}

Inspired by these results, it appears that one can also attempt to unconventionally place the EEZ by choosing an appropriate chain length distribution. One possibility for example would be to place the EEZ away from the substrate, perhaps somewhere in the middle of the polymer brush. To test this idea, we employ the following chain length distribution:
\begin{equation}
\label{eq:mideez }
  P(N)=h_{0} \: \theta(N_{0}-N)
  +h_{1} \: \theta(N-N_{1}) \theta(N_{2}-N),  
\end{equation}
with $N_{0}=0.29$, $N_{1}=0.91$, $N_{2}=1.28$ and the average chain length $N_{a}=1$. $h_{0}$ and $h_{1}$ are chosen so that 10 \% of the chains have chain lengths below $N_{0}$ and the rest of the chain lengths are between $N_{1}$ and $N_{2}$. The rationale behind this choice of chain length distribution is that the ends of chains of size $N_{0}$ would mark the beginning of the EEZ, while the ends of chains of size $N_{1}$ would determine the end of it. We would thus expect an EEZ to occur in some interval $z \in [z_{1},z_{2}]$, which is apriori unknown.
Fig.~\ref{fig:eez_middle} shows that this is indeed the case. We can identify a region where chain ends of shorter chains are located (orange), an EEZ which is devoid of chain ends, and a region containing chain ends of larger chains (blue).

\subsection{\label{sec:chainenddesign}Designing chain end density
profiles}

In practice, another useful quantity one would like to specify instead of the chain length distribution, would be the chain end density. This could be interesting in the case of end-functionalized brush polymers, where the chain end distribution would correspond to the density distribution of functional entities.
By specifying the chain end density and the geometry, it is possible to in return calculate self-consistently, the chain length distribution. The corresponding numerical scheme is given in the Appendix \ref{app:design_chain}.

\begin{figure}
    \centering
    \includegraphics[width=0.48\textwidth]{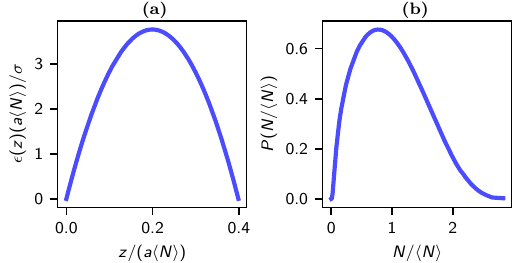}
    \caption{ Projected chain end density versus $z/(a\langle N \rangle)$   on a convex spherical geometry with $R/(aN_{a})=0.1$ and $\sigma w / a$=1 \revv{(a)}. The resulting chain length distribution is shown \revv{in (b)}. 
     }
    \label{fig:from_chend}
\end{figure}

We demonstrate this procedure with the following example projected chain end density:

\begin{eqnarray}
    \epsilon\left((z/aN_{o})(\frac{\sigma w}{a})^{-1/3}\right)&=&\left((z/aN_{o})(\frac{\sigma w}{a})^{-1/3}\right)\\&&\left(1-\left((z/aN_{o})(\frac{\sigma w}{a})^{-1/3}\right)\right),\nonumber
\end{eqnarray}
where $N_{o}$ is a reference chain length and not necessarily the average chain length $\langle N \rangle$.
This projected chain end distribution can be implemented by choosing a chain length distribution as
shown in Fig. \ref{fig:from_chend}.

However, one must note that not all chain end densities can be enforced and that the approximations of SST such as positive tension etc., need to be obeyed for the solution to be physical and to be in accordance with SST.

\section{Conclusion}

We have extended the mean-field theory model of Milner
et.al.\cite{Milner1988,Milner1989}, which employs the
Strong-Stretching limit, to curved substrates, for both convex and
concave geometries for simple polydisperse linear polymers in a good
solvent \revv{and the melt}.

We have \revv{focused} primarily on spherical 
geometries \revv{in good solvent} and investigated the variations in the properties of the brushes composed of experimentally relevant chain length distributions such as the Schulz-Zimm distribution for different polydispersity indices. Among other, we found that in convex geometries, polydispersity has a strong impact on the properties of the chain end distribution as a function of grafting density and geometry, in particular in the range $H \: h \sim 0-10$. In contrast, the influence of polydispersity on the properties of concave brushes is much less pronounced in the regime where \revv{the theory can be applied}.

Furthermore, we have investigated the bending modulus, the Gaussian modulus, and the spontaneous curvature -- which characterizes the driving force towards bending --
as a function of polydispersity and $\sigma w /a$ for surfaces decorated with such polymer brushes. We  found that the moduli are remarkably unaffected by polydispersity. \revv{Interestingly, the (small) impact of polydispersity on the bending moduli for brushes in good solvent is opposite to that for molten brushes: Brushes in good solvent stiffen, whereas molten brushes become softer. On the other hand,}  the analysis of the spontaneous curvature indicates that polydispersity reduces the tendency of brushes to bend \revv{for both types of brushes.} 

Moreover we have successfully accounted for the emergence of End
Exclusion Zones in distributions with non-zero minimum chain length. We have shown that End Exclusion Zones can appear in arbitrary
positions in the brush, as long as the right chain length
distribution, geometry and grafting density are chosen. 
Finally, we have also developed a method for engineering specific chain end distributions by manipulating the chain length distribution and geometry.

Although we have focused on the good solvent case, the same
procedure can be applied to the melt case, as discussed in 
Section \ref{sec:Analytical} (Eq.~(\ref{eq:phi_melt})) and
Appendix \ref{app:free_energy}. In the limiting case of
a uniform distribution with vanishing width, 
\revv{we can compare our numerical results for the EEZ
to those of Dimitriyev {\em et al.} \cite{dimitriyev2021end}.
The agreement is excellent. The comparison can be found in the supporting material.}
\section{Data}
The data shown here along with the code that performs the iteration loops described in Appendix \ref{app:sst_algorithm} and Appendix \ref{app:design_chain} can be found under the link: \url{https://gitlab.rlp.net/mgiannak/polydispersity-in-curved-substrates-sst}.
\begin{acknowledgments}
\revv{The authors are grateful to Le Qiao for valuable discussions and comments.} This work was funded by the German Science Foundation
(DFG) within Grant number 446008821, and the Agence Nationale de La Recherche, France. Partial funding was also received by the DFG within Grant number 429613790.
M.G. is associate member of the integrated graduate school of the collaborative research center TRR 146 ''Multiscale modeling of soft matter systems'', grant number 233630050.
\end{acknowledgments}

\begin{appendix}
\section{Free energy}\label{model1}
\label{app:free_energy}

We begin with recapitulating the formal derivation of the
SCF theory and the resulting SCF free energy. The starting
point is the partition function of the system, 
\begin{equation}\label{eq:partition}
    \mathcal{Z} \propto  \prod_{i}
    \int \ddd \bb{r}_{i} 
    \exp\left[- \frac{\mathcal{H}[\{\bb{r}_i\}]}{k_{B}T} \right]
\end{equation} 
The Hamiltonian ${\mathcal H}$ has been introduced in Eq.~(\ref{eq:ham}). 
The sum $i$ runs over all chains in the system, which are
taken to be distinguishable because they have different
lengths $N_i$. After some field-theoretic transformations\cite{schmid1998self-consistent}, 
the partition function is expressed as an integral over fluctuating
fields $\phi_f$ and $W_f$,
\begin{equation}
\label{eq:partition_field}
    \mathcal{Z}  \propto \int \ddd \phi_f \int_{i \infty} \ddd W_f  \: 
    \exp\left(- \frac{{\mathcal F}[\phi_f,W_f]}{k_B T} \right) 
\end{equation}
with
\begin{equation}
\label{eq:ff}
\frac{{\mathcal F}[\phi_f,W_f]}{k_B T} = 
     {\cal V}[\phi_f]-\int d\bb{r} \, W_f(\bb{r}) \phi_f(\bb{r})
    -\sum_i  \ln (Q_{i}[W_f]).
\end{equation} 
Here $Q_{i}[W_f]$ is the single chain partition 
function of a chain of length $N_i$ in the field $W_f(\bb{r})$,
\begin{equation}
\label{eq:parchain} 
 Q_{i}[W]=\int \ddd \bb{r}_{i}   
   \exp\left(- {\cal S}_i[\bb{r}_i;W_f] \right) 
\end{equation}
with ${\cal S}_i[\bb{r}_i;W_f]$ as defined by Eq.~(\ref{eq:action}) in the
main text.  We note that, in Eq.~(\ref{eq:partition_field}), the field
$W_f(\bb{r})$ is purely imaginary. The SCF approximation consists in
approximating the functional integral in (\ref{eq:partition_field}) by the
extremum of the integrand\cite{schmid1998self-consistent}. This results in the free energy
\begin{equation}
\label{eq:scf_free_energy}
F = - k_B T \: \ln({\cal Z}) \approx {\cal F}[\phi,W],
\end{equation}
where $\phi(\bb{r})$ and $W(\bb{r})$ are the fields that extremize the
functional ${\cal F}$ given by Eq.~(\ref{eq:ff}). At the extremum
(saddle point), the field $W(\bb{r})$ is real, 
$W(\bb{r})= \delta {\cal V}/\delta \phi(\bb{r})$, and
$\phi(\bb{r})=-\sum_i \delta \ln{Q_i}/\delta W(\bb{r})$ is effectively
the density of noninteracting graft chains $i$ in the field $W(\bb{r})$.
For brushes in solvent, the field $W(\bb{r})$ can be calculated from
$W(\bb{r})= \delta {\cal V}/\delta \phi(\bb{r})$. 

The same formalism can also be applied to incompressible melts in a
straightforward manner. The integral $\int \ddd \phi_f$ 
in Eq. (\ref{eq:partition_field}) is then replaced by the 
constraint $\phi_f(\bb{r}) \equiv \bar{\phi}$ everywhere, 
and the field $W(\bb{r})$ which extremizes ${\cal F}$ 
is a Lagrange field that enforces this constraint. The
interaction term ${\cal V}[\phi_f] = {\cal V}[\bar{\phi}]$ 
is a constant and can usually be omitted. 

Now, turning to the SST approximation, the main new aspect compared to
the SCF theory is that we apply a second saddle-point approximation to
evaluate the integral in single chain partition function in
Eq.~(\ref{eq:parchain}). This allows us to derive an explicit
expression for $\ln(Q_i)$:
\begin{eqnarray*}
- \ln (Q_i) &\approx&  \min_{[\bb{r}_i]}{\cal S}[\bb{r}_i;W]
\\ &=&
 \int_0^{N_i} \!\!\!\!
     \ud n \: \left( \frac{3}{2 a^{2}} \big(\frac{\ud
 z_{i}}{\ud n}\big)^{2}+W(z_{i}) \right)
 \\ &=&
 \int_0^{N_i} \!\!\!\! \ud n  \: \frac{3}{a^2} \:
    \left(\frac{\ud z_{i}}{\ud n}\right)^{2}
    + \big(\Umax - E(N_i)\big) \: N_i
 \\ &=&
 \sqrt{\frac{3}{2a^2}} \: 
 \int_0^{E(N_i)} \!\! \ud U' \:
 \frac{\ud z}{\ud U'}\: \frac{E(N_i)+\Umax-2U'}{\sqrt{E(N_i)- U'}}.
\end{eqnarray*}
Here we have used Eqs.\ (\ref{eq:uu},\ref{eq:energy}) in the second
step, and (\ref{eq:tension}, \ref{eq:N_U_abel}) along with the 
identity $N_i = N\big(E(N_i)\big)$ in the third step.  

Based on the above expression for $\ln (Q_i)$, we can now calculate the
last term in Eq.~(\ref{eq:scf_free_energy}, \ref{eq:ff}).  To this
end, we sum over chains $i$ grafted onto a patch with substrate area
$A$, taking into account that $Q_i$ only depends on the chain length
$N_i$ and that chains have the chain length distribution $P(N)$ and
the grafting density $\sigma$.
\begin{eqnarray}
\nonumber
\lefteqn{- \frac{1}{A} \: \sum_i \ln (Q_i) =
\sigma \int \: \ud N \: P(N) \: \ln\big(Q(N)\big)}\quad
\\ &=& \nonumber
\int_0^{\Umax} \!\!\!\!\!\!  \ud U \; 
 \frac{\ud \sigma_c}{\ud U}
\sqrt{\frac{3}{2a^2}}  \int_0^U \ud U'\:
\frac{\ud z}{\ud U'}\: \frac{U+\Umax-2U'}{\sqrt{U-U'}}
\\ &=& \nonumber
\int_0^{\Umax} \!\!\!\!\!\!\! \ud U' \; 
 \frac{\ud z}{\ud U'} \sqrt{\frac{3}{2a^2}}  
\int_{U'}^{\Umax} \!\!\!\!\!\!\! \ud U \; 
  \frac{\ud \sigma_c}{\ud U} \frac{U+\Umax-2U'}{\sqrt{U-U'}}
\\ &=& \nonumber
\int_0^{\Umax} \!\!\!\!\!\!\! \ud U' \; \frac{\ud z}{\ud U'} 
  \left( (\Umax-U') \: \lambda(U')
    - \frac{1}{2} \Lambda(U') \right)
\\ &=& \nonumber
\int_0^{\Umax} \!\!\!\!\!\!\! \ud U' \: \lambda(U')
  \left( \frac{\ud z}{\ud U'} \: (\Umax - U') 
         + \frac{z(U')}{2}  \right)
\end{eqnarray}
In the first step, we have set $U=E(N_i)$ and
used Eq.~(\ref{eq:chainendrel}); in the second step, we 
have exchanged the integration bounds; in the third step,
we have used Eq.~(\ref{eq:l_U_abel}) and introduced
the auxiliary quantity
\begin{displaymath}
  \Lambda(U) = - \sqrt{\frac{6}{a^2}}
   \int_{U'}^{\Umax} \!\!\!\!\!\!\! \ud U\; 
   \frac{\ud \sigma_c}{\ud U} \sqrt{U-U'},
\end{displaymath}
noting $\ud \Lambda/\ud U = \lambda(U)$ with $\Lambda(\Umax) = 0$,
and in the last step, we have performed an integration by parts.

The second term in Eq.~(\ref{eq:scf_free_energy}, \ref{eq:ff})
can be rewritten as
\begin{eqnarray}
\nonumber
\lefteqn{\int \ud \bb{r} \: \phi(\bb{r}) \: W(\bb{r})
= A \: \int_0^\infty \ud z \: g(z) \: \phi(z) \: W(z)} \qquad
\\ && 
= A \:\int_0^{\Umax}\!\!\!\!\!\!\!  \ud U'\;
\frac{\ud z}{\ud U'} \: \lambda(U') \: (\Umax - U'),
\label{eq:WW}
\end{eqnarray}
where we have used Eqs.~(\ref{eq:uu}) and (\ref{eq:l_U_direct})
in the last step. 

Putting everything together, we get the following final result
for the free energy per substrate area in SST approximation:
\begin{equation}
\frac{F}{A \: k_B T} = {\cal V}[\phi] 
 + \frac{1}{2} \int_0^{\Umax} \!\!\!\!\!\! \ud U'\; 
   \lambda(U') \: z(U'),
\end{equation}
which is listed as Eq.~(\ref{eq:free_energy_sst}) in the main
text. For melts, the first term, ${\cal V}[\phi]$ vanishes as
discussed earlier. In good solvent, using Eq.~(\ref{eq:excl})
and $W(\bb{r}) = w \phi(\bb{r})$, it takes the form
\begin{equation}
{\cal V}[\phi] 
  = \frac{1}{A} \frac{w}{2} \int \ud \bb{r} \: \phi(\bb{r})^2 
  = \frac{1}{2A} \int \ud \bb{r} \: \phi(\bb{r}) \: W(\bb{r}),
\end{equation}
which can be simplified in the same way as (\ref{eq:WW}),
leading to Eq.~(\ref{eq:free_energy_solvent}) in the main text.

\section{Derivation of self-consistent SST equations and 
additional auxiliary equations}
\label{app:sst_derivation}

In this section, we derive the Eqs.~(\ref{eq:z_U_abel}),
(\ref{eq:sc_U_abel}), and (\ref{eq:dsc_U_abel}) in the main
text, as well as a few other useful auxiliary equations.

The starting point is Eq.~(\ref{eq:tension}) in the main text, which
gives an expression for the tension in a chain $i$, $\ud z_i/\ud n$,
as a function of the integration constant $E(N_i)$ and the potential
$U(z_i)$ at the position $z_i$. We will use this expression to derive
an relation between $N(U)$ and $z(U)$, and one between  $\lambda(U)$
and $\ud \sigma_c/\ud U$. We recall that $N(U)$ is the length of
chains with end monomers situated at distance $z(U)$ from the
substrate, $\lambda(U)$ is the projected monomer density at $z(U)$,
and $(\ud \sigma_c/\ud U) \: \ud U$ is the number of chains grafted on
a patch of unit area with chain ends located in the interval $[z(U),
z(U+\ud U)]$. 

The derivation of the equation for $N(U)$ is straightforward.
We set $U = E(N_i)$, $U'= U\big(z_i(n)\big)$, and drop indices 
$i$ for simplicity. Then we can write
\begin{eqnarray}
\nonumber
N(U) &=& \int_{0}^{N(U)} \!\!\! \ud n 
= \int_0^U \ud U' \: \bigg/ 
\Big( \frac{\ud U'}{\ud z} \cdot \frac{\ud z}{\ud n} \Big)
\\ &=&
\sqrt{\frac{3}{2a^2}}\int_0^U \ud U' \frac{\ud z}{\ud U' }
\frac{1}{\sqrt{U-U' }}.
\label{eq:N_U_abel}
\end{eqnarray}
We should note that, strictly speaking, Eq.\ (\ref{eq:N_U_abel}) has
no physical meaning for values of $U$ inside an EEZ, as the EEZ 
does not contain any chain ends by definition. However, mathematically, 
the expression (\ref{eq:N_U_abel}) is still perfectly defined, as 
$z(U)$ is defined everywhere. Hence we can use Eq.\
(\ref{eq:N_U_abel}) to extrapolate the function $N(U)$ such that it
is defined in the whole range of $U$.

To obtain an equation for $\lambda(U)$ as a function of 
$\ud \sigma_c(U)/\ud U$, we use the fact that chains of length
$N_i$ with conformation $z_i(n)$ deposit $1\big/\frac{\ud
z_i}{\ud n}\big|_{z_i = z}$ monomers in the interval $[z,z+\ud z]$.  
We sum over all chains with chain ends at positions $z_i(N_i) > z$,
i.e., with characteristic integration constants 
$E(N_i)= U\big(z_i(N_i)\big) > U(z)$
(see Eq.\ (\ref{eq:energy})).  This gives
\begin{eqnarray}
\nonumber
\lambda(U) &=& 
\sum_{i: z_i(N_i) > z(U)} 
  \frac{1}{\ud z_i/\ud n}\bigg|_{z_i(n) = z(U)}
\\ &=&
\sqrt{\frac{3}{2a^2}}\int_U^{\Umax} \!\!\!\!\!\!\! \ud U' \;
  \frac{\ud \sigma_c(U')}{\ud U' } \:
  \frac{1}{\sqrt{U'- U}},
\label{eq:l_U_abel}
\end{eqnarray}
where we have now set $U'=E(N_i)$.

Eq.~(\ref{eq:N_U_abel}) can be reversed via an Abel transform
(see Appendix \ref{app:abel}, using Eq.~(\ref{eq:abel_ap}) and
$z(0)=0$), giving an integral equation for $z(U)$ in terms 
of the function $N(U)$:
\begin{displaymath}
z(U) = \sqrt{\frac{2}{3}} \; \frac{a}{\pi} \:
\int_0^U \ud U' \: N(U') \: 
\frac{1}{\sqrt{U-U' }}.
\end{displaymath}
This equation is listed as Eq.~(\ref{eq:z_U_abel}) in the
main text. Similarly, Eq.~(\ref{eq:l_U_abel}) can be reversed,
giving (using Eq.~(\ref{eq:abel_bp}))
\begin{displaymath}
  \frac{\ud \sigma_c}{\ud U} = \sqrt{\frac{2}{3}} \frac{a}{\pi}
    \left( \frac{\lambda(\Umax)}{\sqrt{\Umax - U}}
    - \int_U^{\Umax} \!\!\!\!\!\!\! \ud U'\;
      \frac{\ud \lambda}{\ud U'}
       \frac{1}{\sqrt{U'- U}} \right),
\end{displaymath}
or, (using Eqs.~(\ref{eq:abel_bp2}) and $\sigma_c(\Umax)=\sigma$)
\begin{displaymath}
\sigma_c(U) = \sigma - \sqrt{\frac{2}{3}} \frac{a}{\pi}
\int_U^{\Umax} \!\!\!\!\! \ud U' \: 
  \frac{\lambda(U')}{\sqrt{U'- U}}.
\end{displaymath}
These equations are listed as Eq.~(\ref{eq:dsc_U_abel}) 
and Eq.~(\ref{eq:sc_U_abel}) in the main text.

For future reference, we also introduce an equation that
expresses $z(U)$ as a function of $\lambda(U)$,
and is obtained directly from (\ref{eq:l_U_direct}):
\begin{equation}
\label{eq:z_U_direct}
z(U) = \left\{\begin{array}{ll}
\frac{H}{K}
\left(-1+\sqrt{1-\frac{K}{H^2}(1-\frac{\lambda(U)}{\phi(U)}} \right)
& \text{for  $K \neq 0$} 
\\
\frac{1}{2 H} \: \left( - 1 + \frac{\lambda(U)}{\phi(U)} \right)
& \text{for  $K = 0$} 
\end{array} \right. 
\end{equation}
This equation cannot be applied for planar surfaces, but we
also do not need it there.

Finally, we note that in numerical implementations, it is convenient
to re-express the integrals in Eqs.\ (\ref{eq:N_U_abel}) and
(\ref{eq:dsc_U_abel}) in terms of integrands that involve $z(U')$ and
$\lambda(U')$ rather than their derivatives. Furthermore, to 
improve the numerical accuracy, we re-arrange the expressions
Eqs.~(\ref{eq:z_U_abel}), (\ref{eq:sc_U_abel}),
(\ref{eq:dsc_U_abel}), and (\ref{eq:N_U_abel}) such that 
integrands do not diverge in the vicinity of integration boundaries. 
This can be achieved by rewriting the equations as
\begin{equation}
 z(U) = \textstyle \sqrt{\frac{2}{3}} \frac{a}{\pi} 
    \Big( 2 N(U) \sqrt{U} 
        - \: \int\limits_0^U \ud U' \: \frac{N(U)-N(U')}{\sqrt{U-U'}}
\Big),
\label{eq:z_U_abel_2}
\end{equation}
\begin{equation}
  \sigma_c(U) = \textstyle \sigma - \sqrt{\frac{2}{3}} \frac{a}{\pi}
    \Big( 2 \lambda(U) \sqrt{\Umax-U} 
        + \: \int\limits_U^{\Umax} \!\!\! \ud U' \:
           \frac{\lambda(U') - \lambda(U)}{\sqrt{U'-U}} \Big),
\label{eq:sc_U_abel_2}
\end{equation}
\begin{equation}
  \lambda(U) = \textstyle \sqrt{\frac{3}{2 a^2}}
    \Big( 2 \sigma_c'(U) \sqrt{\Umax-U}
 + \: \int\limits_U^{\Umax} \!\!\! \ud U' \:
   \frac{\sigma_c'(U')-\sigma_c'(U)}{\sqrt{U' - U}}
\Big)
\label{eq:l_U_abel_2}
\end{equation}
and (using (\ref{eq:abel_ap2}) and (\ref{eq:abel_bp2}))
\begin{eqnarray}
   N(U) &=& \textstyle \sqrt{\frac{3}{2 a^2}} 
    \Big( \frac{z(U)}{\sqrt{U}} + z'(U) \: \sqrt{U} 
\nonumber \\ && \textstyle
        + \: \frac{1}{2} \int\limits_0^U \ud U' \: 
            \frac{z(U)-z(U') - z'(U)\: (U-U' )}{\sqrt{U-U'}^3} 
     \Big),
\label{eq:N_U_abel_2}
\end{eqnarray}
\begin{eqnarray}
\frac{\ud \sigma_c}{\ud U} &=& \textstyle
\sqrt{\frac{2}{3}} \: \frac{a}{\pi} \Big(
\frac{\lambda(U)}{\sqrt{\Umax-U}} - \lambda'(U) \: \sqrt{\Umax-U}
\nonumber \\&& \textstyle
- \: \frac{1}{2} \int\limits_U^{\Umax} \!\!\! \ud U' 
\frac{\lambda(U') - \lambda(U) - \lambda'(U)\: (U' -U)}{\sqrt{U' - U}^3}
\Big).
\label{eq:dsc_U_abel_2}
\end{eqnarray}

\section{Planar and monodisperse brushes}
\label{app:planar_monodisperse}

For planar brushes, we have $g(z) \equiv 1$ in 
Eq.\ (\ref{eq:l_U_direct}) and recover the results of Milner 
{\em et al.} \cite{Milner1989}.  In particular, the integral in 
Eq.~(\ref{eq:sc_U_abel}) can be evaluated explicitly and one obtains
\begin{equation}
\label{eq:sc_planar_solvent}
\sigma_c(U) = \sigma \: (\: 1 - \sqrt{1-U/\Umax}^3 \:)
\end{equation}
for brushes in good solvent and
\begin{equation}
\label{eq:sc_planar_melt}
\sigma_c(U) = \sigma \: (\: 1 - \sqrt{1-U/\Umax} \:)
\end{equation}
for melt brushes. These results are independent of the
chain end distribution. 

We can also consider specifically the monodisperse case
$N(U) \equiv N_{o}$. For planar and concave brushes, where
no EEZ is present, Eq.\ (\ref{eq:z_U_abel}) then gives 
\begin{equation}
\label{eq:z_U_planar}
z(U) = \sqrt{\frac{8}{3}} \frac{a}{\pi} N_{o} \sqrt{U},
\end{equation}
and one recovers the familiar parabolic profile
\cite{Milner1988,zhulina1990theory}
$U(z) = \Umax (z/h)^2$ with $h = z(\Umax)$. 
This result is independent of curvature. For convex
brushes, an EEZ emerges, and chain ends are present
($\ud \sigma_c/\ud U = 0$) in a regime 
$U \in [0:U_{\text{min}}]$. In this regime,
$N(U)$ must be determined self-consistently. It
increases monotonically and can no longer be expressed 
in a simple analytical form \cite{dimitriyev2021end}.

For planar monodisperse brushes, we can combine
Eqs.~(\ref{eq:z_U_planar}) with Eqs.~(\ref{eq:sc_planar_solvent}) 
or (\ref{eq:sc_planar_melt}) and calculate further properties 
of planar brushes such as, e.g., the chain end density:
\begin{equation}
\varepsilon(z) = \frac{\ud \sigma_c}{\ud U} \: \frac{\ud U}{\ud z}
= \sigma \: \frac{z}{h^2}  \: \left\{ \begin{array}{ll}
3 \sqrt{1-(z/h)^2} & \text{solvent}
\\ 1/\sqrt{1-(z/h)^2} & \text{melt}
\end{array}
\right.
\end{equation}

\section{Abel integral equations}
\label{app:abel}

Most integral equations appearing in the SST formalism 
have the form of Abel integrals\cite{abel1881solutions}. For the convenience
of the reader, we recapitulate the relations used in this work.

An integral equation of the form
\begin{displaymath}
f(x) = \int_0^x \ud s \: \frac{\varphi(s)}{\sqrt{x-s}}
\end{displaymath}
can be inverted\cite{abel1881solutions}  according to
\begin{equation}
\label{eq:abel_a}
\varphi(x) = \frac{1}{\pi} 
\bigg( \frac{f(0)}{\sqrt{x}} 
  + \int_0^x \ud \tau \: \frac{f'(\tau)}{\sqrt{x-\tau}} \bigg). 
\end{equation}
Specifically, if $\varphi(s) = \ud \Phi(s)/\ud s$, we have
\begin{equation}
\label{eq:abel_ap}
\Phi(x) = \Phi(0) + \frac{1}{\pi} \int_0^x \ud \tau 
   \frac{f(\tau)}{\sqrt{x-\tau}}.
\end{equation}
Eq.~ (\ref{eq:abel_ap}) can be derived from (\ref{eq:abel_a})
by inserting $\Phi(x) = \Phi(0) + \int_0^x \ud \tau \: \varphi(\tau)$
in (\ref{eq:abel_a}), rearranging double integrals and then
performing a partial integration. In addition, we can then
rewrite $f(x)$ in terms of $\Phi(s)$ as
\begin{equation}
\label{eq:abel_ap2}
f(x) = \frac{\Phi(x) - \Phi(0)}{\sqrt{x}}
+ \frac{1}{2} \int_0^x \frac{\Phi(x) - \Phi(s)}{\sqrt{x-s}^3}.
\end{equation}
This can be seen by considering the last term on the r.h.s.
of the equation and performing a partial integration.

Likewise an integral of the form
\begin{displaymath}
f(x) = \int_x^1 \ud s \: \frac{\varphi(s)}{\sqrt{s-x}},
\end{displaymath}
can be inverted according to
\begin{equation}
\label{eq:abel_b}
\varphi(x) = \frac{1}{\pi} 
\bigg( \frac{f(1)}{\sqrt{1-x}} 
  - \int_x^1 \ud \tau \: \frac{f'(\tau)}{\sqrt{\tau-x}} \bigg).
\end{equation}
For $\varphi(s) = \ud \Phi(s)/\ud s$, we have
\begin{equation}
\label{eq:abel_bp}
\Phi(x) = \Phi(1) - \frac{1}{\pi} \int_x^1 \ud \tau 
   \frac{f(\tau)}{\sqrt{\tau-x}}
\end{equation}
and
\begin{equation}
\label{eq:abel_bp2}
f(x) = \frac{\Phi(1) - \Phi(x)}{\sqrt{1-x}}
+ \frac{1}{2} \int_x^1 \frac{\Phi(s) - \Phi(x)}{\sqrt{s-x}^3}.
\end{equation}

\section{Rescaling the SST equations}
\label{app:scaled}

An inspection of the SST equations
(\ref{eq:z_U_abel})--(\ref{eq:dsc_constraint}) and the
auxiliary equations (\ref{eq:N_U_abel}) -- (\ref{eq:dsc_U_abel_2})
shows that their form is invariant if we rescale all
quantities according to
\begin{align*}
	 & U =\widetilde{U} \: \Umax 
	 & N&=N_{o} \: \widetilde{N}
	 & z&=\widetilde{z}\: \Umax^{\frac{1}{2}} \: a N_{o}
\\ 
     & \phi = \widetilde{\phi} \:  \Umax  / w
     & \omit\rlap{\text{(Melt brushes: Set
     $w:=\Umax/\bar{\phi}_{\text{melt}}$)}}
\\
	 & \sigma = \widetilde{\sigma} \: \frac{\Umax^{\frac{3}{2}}a}{w}
     & \frac{\ud \sigma_c}{\ud U} &= 
       \frac{\ud \widetilde{\sigma_c}}{\ud \widetilde{U}} \:
       \frac{\Umax^{\frac{1}{2}} a}{w}
	 & \lambda&= \widetilde{\lambda} \: \frac{\Umax}{w} 
\\
	 & H = \frac{\widetilde{H}}{a \Umax^{1/2} N_{o}} 
	 & \epsilon(z) &= \widetilde{\epsilon}(\widetilde{z}) \: \frac{\Umax}{w N_{o}}
	 & F &= \widetilde{F} \: \frac{a N_{o} \Umax^{\frac{5}{2}}}{w}
\\& K =\frac{\widetilde{K}}{a^{2}\Umax N_{o}^{2}}
\end{align*}
Here $N_{o}$ is an (arbitrary) reference chain length. Written in these
rescaled quantities, the parameters $a, w$ and $\Umax$ are replaced by
unity in the equations. This means that, when solving the equations, it
is sufficient to consider the case $\Umax = 1, \: a=1, \: w=1, \:
N_o=1$. The solutions for other values of, e.g., $\Umax$ and $N_o$,
can be recovered by rescaling the results accordingly. We note that
the grafting density, $\sigma$, is not an input parameter in the
equations. Rather, $\Umax$ has to be tuned such that the desired 
value of $\sigma$ is obtained.

\section{Numerical Scheme}
\label{app:sst_algorithm}
\begin{figure}
\centering
\begin{tikzpicture}[x=0.75pt,y=0.75pt,yscale=-0.9,xscale=0.85]

\draw  [line width=1.5]  (51.15,61) -- (196.98,61) -- (196.98,136.26) -- (51.15,136.26) -- cycle ;
\draw [line width=2.25]    (196.82,98.17) -- (248.91,98.49) ;
\draw [shift={(253.91,98.52)}, rotate = 180.35] [fill={rgb, 255:red, 0; green, 0; blue, 0 }  ][line width=0.08]  [draw opacity=0] (14.29,-6.86) -- (0,0) -- (14.29,6.86) -- cycle    ;
\draw [line width=2.25]    (327.03,136.48) -- (327.34,188.83) ;
\draw [shift={(327.37,193.83)}, rotate = 269.66] [fill={rgb, 255:red, 0; green, 0; blue, 0 }  ][line width=0.08]  [draw opacity=0] (14.29,-6.86) -- (0,0) -- (14.29,6.86) -- cycle    ;
\draw  [line width=1.5]  (254.29,61) -- (400.11,61) -- (400.11,136.26) -- (254.29,136.26) -- cycle ;
\draw  [line width=1.5]  (254.29,194.43) -- (400.11,194.43) -- (400.11,269.69) -- (254.29,269.69) -- cycle ;
\draw [line width=2.25]    (327.03,268.96) -- (327.34,321.3) ;
\draw [shift={(327.37,326.3)}, rotate = 269.66] [fill={rgb, 255:red, 0; green, 0; blue, 0 }  ][line width=0.08]  [draw opacity=0] (14.29,-6.86) -- (0,0) -- (14.29,6.86) -- cycle    ;
\draw  [line width=1.5]  (254.29,326.91) -- (400.11,326.91) -- (400.11,402.17) -- (254.29,402.17) -- cycle ;
\draw [line width=2.25]    (327.03,401.44) -- (327.34,453.78) ;
\draw [shift={(327.37,458.78)}, rotate = 269.66] [fill={rgb, 255:red, 0; green, 0; blue, 0 }  ][line width=0.08]  [draw opacity=0] (14.29,-6.86) -- (0,0) -- (14.29,6.86) -- cycle    ;
\draw  [line width=1.5]  (254.29,459.82) -- (400.11,459.82) -- (400.11,536.45) -- (254.29,536.45) -- cycle ;
\draw [line width=2.25]    (200.64,457.7) -- (254.38,402.17) ;
\draw [shift={(197.16,461.29)}, rotate = 314.06] [fill={rgb, 255:red, 0; green, 0; blue, 0 }  ][line width=0.08]  [draw opacity=0] (14.29,-6.86) -- (0,0) -- (14.29,6.86) -- cycle    ;
\draw  [line width=1.5]  (51.15,461.29) -- (196.98,461.29) -- (196.98,536.55) -- (51.15,536.55) -- cycle ;
\draw  [line width=1.5]  (51.15,327.86) -- (196.98,327.86) -- (196.98,403.12) -- (51.15,403.12) -- cycle ;
\draw [line width=2.25]    (123.92,408.34) -- (124.23,460.69) ;
\draw [shift={(123.89,403.34)}, rotate = 89.66] [fill={rgb, 255:red, 0; green, 0; blue, 0 }  ][line width=0.08]  [draw opacity=0] (14.29,-6.86) -- (0,0) -- (14.29,6.86) -- cycle    ;
\draw  [line width=1.5]  (51.15,196.88) -- (196.98,196.88) -- (196.98,272.14) -- (51.15,272.14) -- cycle ;
\draw [line width=2.25]    (123.92,277.72) -- (124.23,328.07) ;
\draw [shift={(123.89,272.72)}, rotate = 89.65] [fill={rgb, 255:red, 0; green, 0; blue, 0 }  ][line width=0.08]  [draw opacity=0] (14.29,-6.86) -- (0,0) -- (14.29,6.86) -- cycle    ;
\draw [line width=2.25]    (327.03,536.01) -- (327.34,588.36) ;
\draw [shift={(327.37,593.36)}, rotate = 269.66] [fill={rgb, 255:red, 0; green, 0; blue, 0 }  ][line width=0.08]  [draw opacity=0] (14.29,-6.86) -- (0,0) -- (14.29,6.86) -- cycle    ;
\draw  [line width=1.5]  (254.29,593.96) -- (400.11,593.96) -- (400.11,669.22) -- (254.29,669.22) -- cycle ;
\draw  [line width=1.5]  (147.69,727.74) -- (293.52,727.74) -- (293.52,803) -- (147.69,803) -- cycle ;
\draw [line width=2.25]    (326.5,669.5) -- (295.98,723.39) ;
\draw [shift={(293.52,727.74)}, rotate = 299.53] [fill={rgb, 255:red, 0; green, 0; blue, 0 }  ][line width=0.08]  [draw opacity=0] (14.29,-6.86) -- (0,0) -- (14.29,6.86) -- cycle    ;
\draw  [line width=1.5]  (51.15,595.17) -- (196.98,595.17) -- (196.98,670.43) -- (51.15,670.43) -- cycle ;
\draw [line width=2.25]    (123.92,542.17) -- (124.23,594.52) ;
\draw [shift={(123.89,537.17)}, rotate = 89.66] [fill={rgb, 255:red, 0; green, 0; blue, 0 }  ][line width=0.08]  [draw opacity=0] (14.29,-6.86) -- (0,0) -- (14.29,6.86) -- cycle    ;
\draw [line width=2.25]    (125.66,675.26) -- (147.69,727.74) ;
\draw [shift={(123.73,670.65)}, rotate = 67.23] [fill={rgb, 255:red, 0; green, 0; blue, 0 }  ][line width=0.08]  [draw opacity=0] (14.29,-6.86) -- (0,0) -- (14.29,6.86) -- cycle    ;
\draw [line width=2.25]    (250.86,139.89) -- (196.98,196.88) ;
\draw [shift={(254.29,136.26)}, rotate = 133.39] [fill={rgb, 255:red, 0; green, 0; blue, 0 }  ][line width=0.08]  [draw opacity=0] (14.29,-6.86) -- (0,0) -- (14.29,6.86) -- cycle    ;

\draw (82.56,81.63) node [anchor=north west][inner sep=0.75pt]  [font=\small] [align=left] {Initial guess\\ \ \ \ for $\widetilde{N}(\widetilde{U})$};
\draw (262.2,72.63) node [anchor=north west][inner sep=0.75pt]  [font=\small] [align=left] {\begin{minipage}[lt]{80.27pt}\setlength\topsep{0pt}
\begin{center}
{ Calculate $\widetilde{z}(\widetilde{U})$ via \\ Eq.~(\ref{eq:z_U_abel}) 
     for all  \\the domain of U }
\end{center}

\end{minipage}};
\draw (257.2,210.06) node [anchor=north west][inner sep=0.75pt]  [font=\small] [align=left] {\begin{minipage}[lt]{86.89pt}\setlength\topsep{0pt}
\begin{center}
 { Calculate $\widetilde{\lambda}(\widetilde{U})$ 
     from \\ $\widetilde{z}(\widetilde{U})$ via Eq.~(\ref{eq:l_U_direct}) }
\end{center}

\end{minipage}};
\draw (263.2,342.04) node [anchor=north west][inner sep=0.75pt]  [font=\small] [align=left] {\begin{minipage}[lt]{81.3pt}\setlength\topsep{0pt}
\begin{center}
{Calculate $\ud\widetilde{\sigma}_c(\widetilde{U})/\ud \widetilde{U}$ \\ via Eq.~(\ref{eq:dsc_U_abel}) }
\end{center}

\end{minipage}};
\draw (330.33,409.94) node [anchor=north west][inner sep=0.75pt]  [font=\small,rotate=-359.38] [align=left] {\begin{minipage}[lt]{65.22pt}\setlength\topsep{0pt}
\begin{center}
{ If $\ud\widetilde{\sigma}_c(\widetilde{U})/\ud \widetilde{U}<0$  \\ \ \ for some $\widetilde{U}$}
\end{center}

\end{minipage}};
\draw (255.7,461.13) node [anchor=north west][inner sep=0.75pt]  [font=\small] [align=left] {\begin{minipage}[lt]{90.24pt}\setlength\topsep{0pt}
\begin{center}
 { Set $\ud\widetilde{\sigma}_c(\widetilde{U})/\ud \widetilde{U}<0$ \\to 0, rescale $\widetilde{\lambda}(\widetilde{U})$ \\and $\ud\widetilde{\sigma}_c(\widetilde{U})/\ud \widetilde{U}$   \\ so that 
 $\widetilde{\lambda}(0)=1$ }
\end{center}

\end{minipage}};
\draw (57.56,472.42) node [anchor=north west][inner sep=0.75pt]  [font=\small] [align=left] {\begin{minipage}[lt]{83.33pt}\setlength\topsep{0pt}
\begin{center}
 { Calculate $\widetilde{\sigma}_c(\widetilde{U})$ by \\integrating over \\   $\ud\widetilde{\sigma}_c(\widetilde{U})/\ud \widetilde{U}$  }
\end{center}

\end{minipage}};
\draw (50.06,340.49) node [anchor=north west][inner sep=0.75pt]  [font=\small] [align=left] {\begin{minipage}[lt]{92.52pt}\setlength\topsep{0pt}
\begin{center}
{Apply Eq.~(\ref{eq:N_U_direct}) to \\
      calculate $\widetilde{N}(\widetilde{U})$ outside of the EEZ}
\end{center}

\end{minipage}};
\draw (53.56,205.01) node [anchor=north west][inner sep=0.75pt]  [font=\small] [align=left] {\begin{minipage}[lt]{88.94pt}\setlength\topsep{0pt}
\begin{center}
{Mix new and old \\values of $\widetilde{N}(\widetilde{U})$ for \\ all $\widetilde{U}$ until  ${\cal E} < 10^{-9} $ \\in Eq. (\ref{eq:accuracy})}
\end{center}

\end{minipage}};
\draw (155.7,440.9) node [anchor=north west][inner sep=0.75pt]  [font=\small,rotate=-315.26] [align=left] {If $\ud\widetilde{\sigma}_c(\widetilde{U})/\ud \widetilde{U}>0$  \\ \ \ for all of $\widetilde{U}$};
\draw (251.7,606.59) node [anchor=north west][inner sep=0.75pt]  [font=\small] [align=left] {\begin{minipage}[lt]{96.09pt}\setlength\topsep{0pt}
\begin{center}
{Recalculate $\widetilde{\lambda}(\widetilde{U})$\\in the EEZ \\via Eq. (\ref{eq:l_U_abel_2})}
\end{center}

\end{minipage}};
\draw (150.6,748.37) node [anchor=north west][inner sep=0.75pt]  [font=\small] [align=left] {\begin{minipage}[lt]{89.96pt}\setlength\topsep{0pt}
\begin{center}
{Evaluate $\widetilde{z}(\widetilde{U})$\\in EEZ via Eq. (\ref{eq:z_U_direct})}
\end{center}

\end{minipage}};
\draw (60.06,608.8) node [anchor=north west][inner sep=0.75pt]  [font=\small] [align=left] {\begin{minipage}[lt]{80.28pt}\setlength\topsep{0pt}
\begin{center}
 { Calculate $\widetilde{N}(\widetilde{U})$\\ in the EEZ \\ via Eq. (\ref{eq:z_U_abel_2}) }
\end{center}

\end{minipage}};
\draw (180.48,181.84) node [anchor=north west][inner sep=0.75pt]  [font=\small,rotate=-315.26] [align=left] {If ${\cal E} > 10^{-9} $ \\};

\end{tikzpicture}\caption{Numerical scheme used to solve the self-consistent equations given a chain length distribution. }
\label{fig:flow_chart1}
\end{figure}
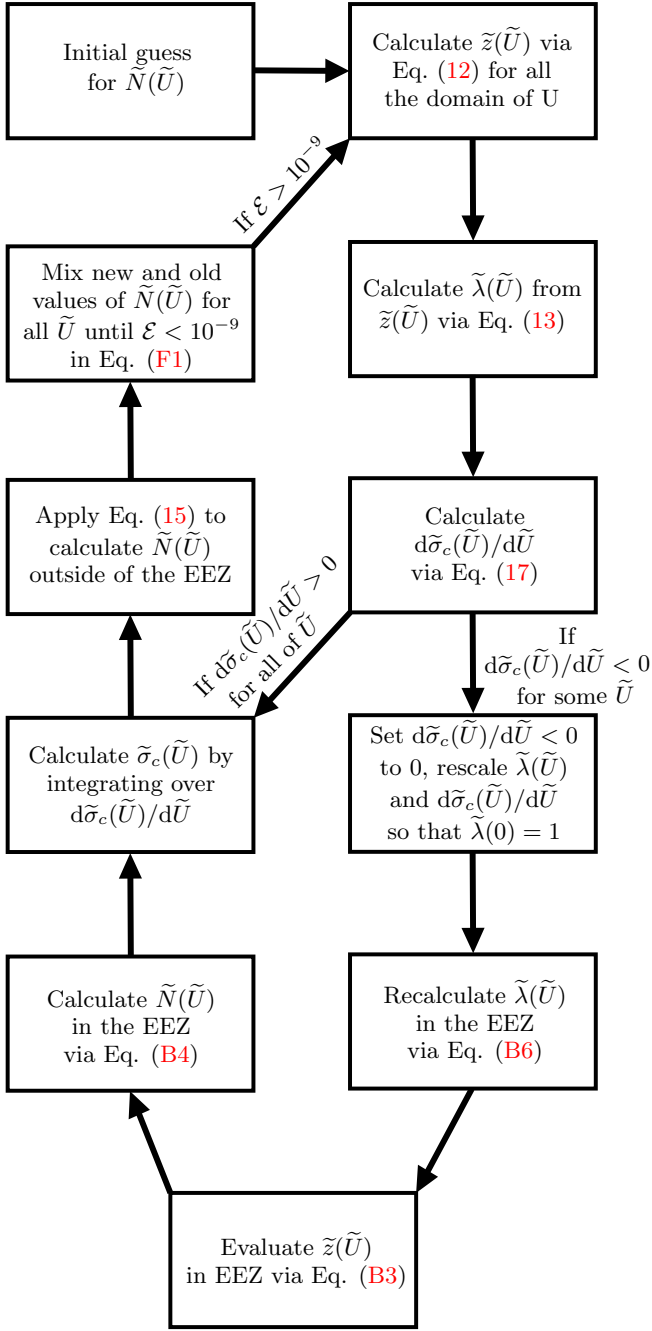

Here we demonstrate our numerical scheme for the solution of the SST
equations given a chain length distribution and geometry $\widetilde{H}$ and $\widetilde{K}$ with the flow chart shown in Fig.~\ref{fig:flow_chart1}. We use the rescaled equations introduced in
Appendix \ref{app:scaled}.

The mixing of new and old values of $\widetilde{N}(\widetilde{U})$ is done via the $\lambda$ mixing method described in Refs.\  \cite{schmid1995quantitative,mueller2007incorporating}  . The value of ${\cal E}$ is defined as:
 \begin{equation}
 \label{eq:accuracy}
    {\cal E} = \int_0^1 \ud \widetilde{U} \: 
   \Big(\widetilde{N}_{new}(\widetilde{U})
        -\widetilde{N}_{old}(\widetilde{U})\Big)^{2}.
\end{equation}

\tikzset{every picture/.style={line width=0.75pt}} 

We conclude with a few comments on numerical aspects of
the algorithm:
\begin{itemize}
\item In practice, it is convenient to use Eqs.~(\ref{eq:z_U_abel_2})-(\ref{eq:dsc_U_abel_2}) for evaluating the Abel integrals.
 Nevertheless, the numerical (discretization) errors of these integrals 
 may cause problems which can be reduced by some tricks.

\item
An additional check is that in the converged solution, the number of monomers above a unit area should be equal to:
 \begin{equation*}
      \widetilde{\sigma} \:  \int_{0}^{\widetilde{N}_{max}}
      \ud\widetilde{N}' \: \widetilde{N}' \: 
      \widetilde{P}(\widetilde{N}')=\widetilde{\sigma} \langle
      \widetilde{N}\rangle=\int_{0}^{\widetilde{h}} d\widetilde{z} \,
      \widetilde{\lambda}(\widetilde{U}).
\end{equation*}
\end{itemize}

\section{Scheme for designing chain end densities}\label{app:design_chain}

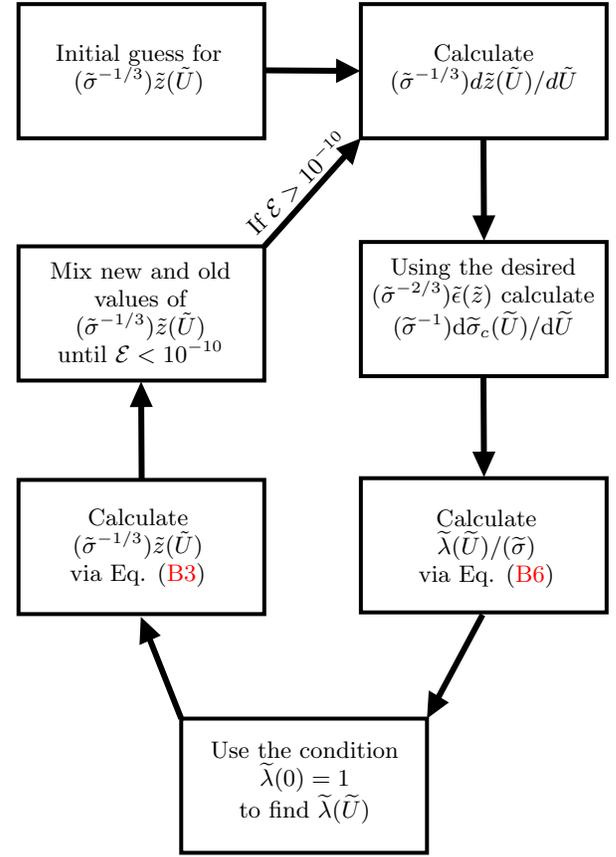
\begin{figure}
    \centering
    \begin{tikzpicture}[x=0.75pt,y=0.75pt,yscale=-0.9,xscale=0.85]

\draw  [line width=1.5]  (51.15,61) -- (196.98,61) -- (196.98,136.26) -- (51.15,136.26) -- cycle ;
\draw [line width=2.25]    (196.82,98.17) -- (248.91,98.49) ;
\draw [shift={(253.91,98.52)}, rotate = 180.35] [fill={rgb, 255:red, 0; green, 0; blue, 0 }  ][line width=0.08]  [draw opacity=0] (14.29,-6.86) -- (0,0) -- (14.29,6.86) -- cycle    ;
\draw [line width=2.25]    (327.03,136.48) -- (327.34,188.83) ;
\draw [shift={(327.37,193.83)}, rotate = 269.66] [fill={rgb, 255:red, 0; green, 0; blue, 0 }  ][line width=0.08]  [draw opacity=0] (14.29,-6.86) -- (0,0) -- (14.29,6.86) -- cycle    ;
\draw  [line width=1.5]  (254.29,61) -- (400.11,61) -- (400.11,136.26) -- (254.29,136.26) -- cycle ;
\draw  [line width=1.5]  (254.29,194.43) -- (400.11,194.43) -- (400.11,269.69) -- (254.29,269.69) -- cycle ;
\draw [line width=2.25]    (327.03,268.96) -- (327.34,321.3) ;
\draw [shift={(327.37,326.3)}, rotate = 269.66] [fill={rgb, 255:red, 0; green, 0; blue, 0 }  ][line width=0.08]  [draw opacity=0] (14.29,-6.86) -- (0,0) -- (14.29,6.86) -- cycle    ;
\draw  [line width=1.5]  (254.29,326.91) -- (400.11,326.91) -- (400.11,402.17) -- (254.29,402.17) -- cycle ;
\draw  [line width=1.5]  (51.15,327.86) -- (196.98,327.86) -- (196.98,403.12) -- (51.15,403.12) -- cycle ;
\draw  [line width=1.5]  (51.15,196.88) -- (196.98,196.88) -- (196.98,272.14) -- (51.15,272.14) -- cycle ;
\draw [line width=2.25]    (123.92,277.72) -- (124.23,328.07) ;
\draw [shift={(123.89,272.72)}, rotate = 89.65] [fill={rgb, 255:red, 0; green, 0; blue, 0 }  ][line width=0.08]  [draw opacity=0] (14.29,-6.86) -- (0,0) -- (14.29,6.86) -- cycle    ;
\draw  [line width=1.5]  (147.69,461.74) -- (293.52,461.74) -- (293.52,537) -- (147.69,537) -- cycle ;
\draw [line width=2.25]    (326.5,403.5) -- (295.98,457.39) ;
\draw [shift={(293.52,461.74)}, rotate = 299.53] [fill={rgb, 255:red, 0; green, 0; blue, 0 }  ][line width=0.08]  [draw opacity=0] (14.29,-6.86) -- (0,0) -- (14.29,6.86) -- cycle    ;
\draw [line width=2.25]    (125.66,409.26) -- (147.69,461.74) ;
\draw [shift={(123.73,404.65)}, rotate = 67.23] [fill={rgb, 255:red, 0; green, 0; blue, 0 }  ][line width=0.08]  [draw opacity=0] (14.29,-6.86) -- (0,0) -- (14.29,6.86) -- cycle    ;
\draw [line width=2.25]    (250.86,139.89) -- (196.98,196.88) ;
\draw [shift={(254.29,136.26)}, rotate = 133.39] [fill={rgb, 255:red, 0; green, 0; blue, 0 }  ][line width=0.08]  [draw opacity=0] (14.29,-6.86) -- (0,0) -- (14.29,6.86) -- cycle    ;

\draw (60.06,82.13) node [anchor=north west][inner sep=0.75pt]  [font=\small] [align=left] {\begin{minipage}[lt]{77.44pt}\setlength\topsep{0pt}
\begin{center}
Initial guess for \\$(\Tilde{\sigma}^{-1/3}) \Tilde{z}(\Tilde{U})$
\end{center}

\end{minipage}};
\draw (255.7,82.13) node [anchor=north west][inner sep=0.75pt]  [font=\small] [align=left] {\begin{minipage}[lt]{88.98pt}\setlength\topsep{0pt}
\begin{center}
 Calculate $(\Tilde{\sigma}^{-1/3})d\Tilde{z}(\Tilde{U})/d\Tilde{U}$ 
\end{center}

\end{minipage}};
\draw (249.7,202.06) node [anchor=north west][inner sep=0.75pt]  [font=\small] [align=left] {\begin{minipage}[lt]{96.15pt}\setlength\topsep{0pt}
\begin{center}
 Using the desired \\$(\Tilde{\sigma}^{-2/3})\Tilde{\epsilon}(\Tilde{z})$ calculate $(\widetilde{\sigma}^{-1})\ud\widetilde{\sigma}_c(\widetilde{U})/\ud \widetilde{U}$
\end{center}

\end{minipage}};
\draw (255.2,343.04) node [anchor=north west][inner sep=0.75pt]  [font=\small] [align=left] {\begin{minipage}[lt]{91.66pt}\setlength\topsep{0pt}
\begin{center}
 Calculate\\   $\widetilde{\lambda}(\widetilde{U})/(\widetilde{\sigma})$ \\via Eq. (\ref{eq:l_U_abel_2}) 
\end{center}

\end{minipage}};
\draw (144.1,472.87) node [anchor=north west][inner sep=0.75pt]  [font=\small] [align=left] {\begin{minipage}[lt]{95.77pt}\setlength\topsep{0pt}
\begin{center}
 Use the condition  $\widetilde{\lambda}(0)=1$\\ to find $\widetilde{\lambda}(\widetilde{U})$
\end{center}

\end{minipage}};
\draw (65.56,204.01) node [anchor=north west][inner sep=0.75pt]  [font=\small] [align=left] {\begin{minipage}[lt]{71.65pt}\setlength\topsep{0pt}
\begin{center}
Mix new and old \\values of $(\Tilde{\sigma}^{-1/3}) \Tilde{z}(\Tilde{U})$ \\until ${\cal E} < 10^{-10}$
\end{center}

\end{minipage}};
\draw (67.56,342.49) node [anchor=north west][inner sep=0.75pt]  [font=\small] [align=left] {\begin{minipage}[lt]{67.23pt}\setlength\topsep{0pt}
\begin{center}
Calculate $(\Tilde{\sigma}^{-1/3}) \Tilde{z}(\Tilde{U})$\\via Eq. (\ref{eq:z_U_direct})
\end{center}

\end{minipage}};
\draw (181.48,182.84) node [anchor=north west][inner sep=0.75pt]  [font=\small,rotate=-315.26] [align=left] {If ${\cal E} > 10^{-10}$ \\};
\end{tikzpicture}
    \caption{Numerical scheme used to solve the self-consistent equations given a projected chain end density.}
    \label{fig:flow_chart2}
\end{figure}

Here we demonstrate our numerical scheme for solving the SST
equations given a chain end density and geometry $\widetilde{H}\widetilde{\sigma}^{1/3}$ and $\widetilde{K}\widetilde{\sigma}^{2/3}$ with the flow chart shown in Fig.~\ref{fig:flow_chart2}. 

In this case we define:
\begin{equation}
 \label{eq:accuracy1}
    {\cal E} = (\Tilde{\sigma}^{-1/3})\int_0^1 \ud \widetilde{U} \: 
     \Big(\widetilde{z}_{new}(\widetilde{U})
        -\widetilde{z}_{old}(\widetilde{U})\Big)^{2}.
\end{equation}

\end{appendix}
\bibliography{brush}
\clearpage

\renewcommand{\thetable}{S\arabic{table}}
\renewcommand{\thefigure}{S\arabic{figure}}
\renewcommand{\thesection}{S\arabic{section}}
\renewcommand{\theHtable}{S\arabic{table}}
\renewcommand{\theHfigure}{S\arabic{figure}}
\renewcommand{\theHsection}{S\arabic{section}}
\setcounter{figure}{0}
\setcounter{section}{0}
\section*{Supporting Material}
Here, we show a comparison of our results for the EEZ with results from Dimitriyev {\em et al.} \cite{dimitriyev2021end} and two examples of shapes of ''potentials'' $U(z)$ inside the EEZ of brushes.
\vspace{10mm}
\section{Test of EEZ results against literature data}
\begin{figure}[H]
    \centering
    \includegraphics[width=0.48\textwidth]{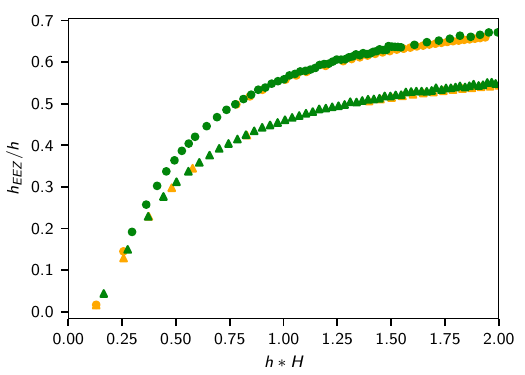}
    \caption{$h*H$ against $h_{eez}/h$ of our results (orange) against Dimitriyev et al. (green) results for both spherical ($\bullet$) and cylindrical ($\blacktriangle$) geometries. 
    These calculations were performed for collapsed melt brushes (with constant density), and not for brushes in good solvent as in the main text. The monodisperse brush was approximated by a brush with very narrow uniform distribution, i.e. $N_{min}=0.9999$ and $N_{max}=1.0001$.}
    \label{fig:melt_comp}
\end{figure}

To validate our SST approach, we tested it by comparing
the results from SST calculations using our approach
with published data from \cite{dimitriyev2021end} where the authors
explored the melt case in the monodisperse limit in various geometries. We do this by taking the limit of decreasing width for the uniform distribution to approximate the case of the monodisperse limit. We plot the results for the ratio between the EEZ thickness and the thickness of the brush versus the product of the thickness of the brush and the mean curvature Fig.~\ref{fig:melt_comp}. There appears to be good agreement.
\section{Potentials in EEZ}
\begin{figure}[H]
    \centering
    \includegraphics[width=0.48\textwidth]{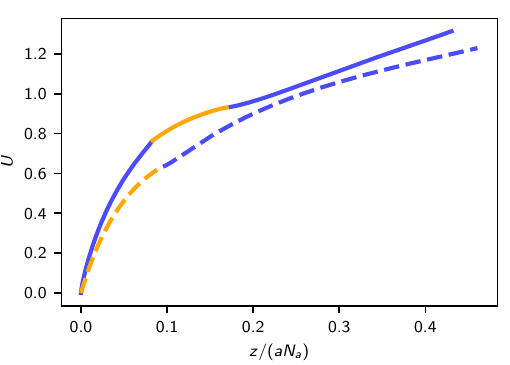}
    \caption{Plots of the potentials $U$ against the distance from the substrate for spherically convex geometries with $\sigma w /a$=1 and $R/(aN_{a})$=0.1 for two different chain length distributions. The 'solid' line represents the 'double step' distribution investigated in the main text, while the 'dashed' line represents the 'single step' distribution with polydispersity index of $N_{w}/N_{a}$=1.1. The 'blue' and 'orange' regions in each plot, represent the potential for the brush outside and inside of the EEZ respectively.}
    \label{fig:potential}
\end{figure}

In Fig.~\ref{fig:potential} we plot the potential against the distance for two different chain length distributions, where the EEZ is either adjacent or not adjacent to the substrate. As it can be seen, the potentials still monotonically increase with $z$ as per the assumptions made in the model. 
\end{document}